\documentclass[preprint]{aastex}

\usepackage{graphicx}
\usepackage{amsmath}
\usepackage{ulem}

\usepackage{natbib}

\newcommand{\ex}{\hat{\boldsymbol e}_x}
\newcommand{\ey}{\hat{\boldsymbol e}_y}
\newcommand{\ez}{\hat{\boldsymbol e}_z}

\newcommand{\rhoi}{\rho_i}
\newcommand{\rhoe}{\rho_e}

\newcommand{\vai}{v_{Ai}}
\newcommand{\vae}{v_{Ae}}
\newcommand{\vaie}{v_{Ai,e}}
\newcommand{\va}{v_{A}}

\newcommand{\tauai}{\tau_{Ai}}

\newcommand{\rhoO}{\rho_0}
\newcommand{\BO}{B_0}
\newcommand{\vecBO}{{\boldsymbol B}_0}

\newcommand{\vecV}{{\boldsymbol V}}
\newcommand{\vecB}{{\boldsymbol B}}

\newcommand{\vecv}{\boldsymbol{v}}
\newcommand{\vx}{v_x}
\newcommand{\vy}{v_y}
\newcommand{\vz}{v_z}

\newcommand{\vecb}{\boldsymbol{b}}
\newcommand{\bx}{b_x}
\newcommand{\by}{b_y}
\newcommand{\bz}{b_z}

\newcommand{\hvx}{\hat{v}_x}
\newcommand{\hvy}{\hat{v}_y}
\newcommand{\hbx}{\hat{b}_x}
\newcommand{\hby}{\hat{b}_y}
\newcommand{\hbz}{\hat{b}_z}

\newcommand{\mi}{m_i}
\newcommand{\me}{m_e}
\newcommand{\mie}{m_{i,e}}

\newcommand{\kappai}{\kappa_i}
\newcommand{\kappae}{\kappa_e}
\newcommand{\kappaie}{\kappa_{i,e}}

\newcommand{\Lx}{L_x}

\newcommand{\vO}{v_0}

\newcommand{\Nx}{N_x}
\newcommand{\Nz}{N_z}

\newcommand{\vxceof}{\tilde v_x}
\newcommand{\Svx}{S_{v_x}}
\newcommand{\thetavx}{\theta_{v_x}}
\newcommand{\vyceof}{\tilde v_y}
\newcommand{\Sivy}{S_{iv_y}}
\newcommand{\thetaivy}{\theta_{iv_y}}
\newcommand{\bzceof}{\tilde b_z}
\newcommand{\Sbz}{S_{b_z}}
\newcommand{\thetabz}{\theta_{b_z}}

\newcommand{\Dvx}{\Delta v_x}
\newcommand{\Dvy}{\Delta v_y}
\newcommand{\Dbz}{\Delta b_z}

\newcommand{\epsvx}{\varepsilon_{v_x}}
\newcommand{\epsvy}{\varepsilon_{iv_y}}
\newcommand{\epsbz}{\varepsilon_{ib_z}}

\newcommand{\difvx}{\delta_{v_x}}
\newcommand{\difvy}{\delta_{iv_y}}
\newcommand{\difbz}{\delta_{ib_z}}

\shorttitle{Normal modes of transverse coronal loop oscillations from numerical simulations: I. Method and test case}
\shortauthors{Rial et al.}

\begin{document}

\title{Normal modes of transverse coronal loop oscillations from numerical simulations: I. Method and test case}

\author{S. Rial\altaffilmark{1}, I. Arregui\altaffilmark{2}, R. Oliver\altaffilmark{1,3}, and J. Terradas\altaffilmark{1,3}}
\altaffiltext{1}{Departament de F\'\i sica, Universitat de les Illes Balears, 07122 Palma de Mallorca, Spain}
\altaffiltext{2}{Instituto de Astrof\'\i sica de Canarias, 38205 La Laguna, Spain}
\altaffiltext{3}{Institute of Applied Computing \& Community Code (IAC3), UIB, Spain}
\email{ramon.oliver@uib.es}

\begin{abstract}
The purpose of this work is to develop a procedure to obtain the normal modes of a coronal loop from time-dependent numerical simulations with the aim of better understanding observed transverse loop oscillations. To achieve this goal, in this paper we present a new method and test its performance with a problem for which the normal modes can be computed analytically. In a follow-up paper, the application to the simulations of \citet{rial2013} is tackled.
The method proceeds iteratively and at each step consists of (i) a time-dependent numerical simulation followed by (ii) the Complex Empirical Orthogonal Function (CEOF) analysis of the simulation results. The CEOF analysis provides an approximation to the normal mode eigenfunctions that can be used to set up the initial conditions for the numerical simulation of the following iteration, in which an improved normal mode approximation is obtained. The iterative process is stopped once the global difference between successive approximate eigenfunctions is below a prescribed threshold.
The equilibrium used in this paper contains material discontinuities that result in one eigenfunction with a jump across these discontinuities and two eigenfunctions whose normal derivatives are discontinuous there. After 6 iterations, the approximation to the frequency and eigenfunctions are accurate to $\lesssim 0.7$\% except for the eigenfunction with discontinuities, which displays a much larger error at these positions.
\end{abstract}


\section{INTRODUCTION}
\label{sec:intro}


The solar atmosphere is the site of a diversity of magnetohydrodynamic waves and oscillations. Transverse coronal loop oscillations are a prominent example of such events. They take place when a large energy deposition, usually caused by a flare, perturbs an active region magnetic structure, which sets some loops into oscillation \citep[see, e.g.,][for some early observations]{aschwanden1999,nakariakov1999}. These events have been modeled with the help of slab and straight cylindrical loop models, whose normal modes can often be obtained by either analytical or numerical means \citep[see][for a review]{ruderman2009}. Starting with the simplest model that considers the fundamental transverse oscillation of a magnetic flux tube \citep{roberts1981,edwin1983,nakariakov2005} several model improvements have included other effects, such as the curvature of coronal loops \citep{vandoorsselaere2004,vandoorsselaere2009,terradas2006}, longitudinal density stratification \citep{andries2005a,andries2005b}, magnetic field expansion \citep{ruderman2008}, departure from circular cross section of the tubes \citep{ruderman2003}, or coronal loop cooling \citep{aschwanden2008,morton2009}. These ingredients have been seen to produce effects on the main wave properties, such as shifts on the frequency and position of the antinodes of the eigenfunctions. Also, the presence of internal fine structuring \citep{terradas2008} and/or a continuous cross-field inhomogenity in density is known to produce important effects, making possible physical processes such as phase-mixing \citep{heyvaerts1983} and resonant damping \citep{hollweg1988,ruderman2002,goossens2002}. The more general the model, the more difficult it is to calculate the eigenmodes of the structure and one has to resort to the study of time-dependent numerical simulations to study these transverse oscillations \citep{selwa2006,selwa2007,selwa2011a,selwa2011b,rial2013}. However, the comparison between the obtained numerical results to observed properties is not as straightforward as with the use of simple models. In these simulations, the initial disturbance excites different oscillatory harmonics, whose presence in the results is easily detected by a Fourier analysis of the variables collected at different points in the numerical domain, but this does not give information about the spatial structure of the eigenmodes. Hence, direct comparison between observed wave properties and the possibly present normal modes becomes difficult. For this reason, we have decided to devise the algorithm described in this paper, which allows us to isolate the eigenmodes present in a numerical simulation. Given the space required to present the algorithm, its application to the time-dependent numerical simulations of \citet{rial2013}, who use a model that takes into account effects such as density stratification, curvature, etc., is left for the second part of this work \citep{rial2019}.


Normal modes provide a physical basis to understand the dynamics of a system. When the equilibrium configuration does not allow a simple solution of the normal mode problem, numerical techniques must be used to determine the normal modes' eigenfunctions and eigenfrequencies. However, general purpose (i.e., for arbitrary equilibria) numerical codes that provide this information cannot be readily found. On the other hand, general purpose numerical codes to solve time-dependent equations are much more abundant. For this reason, being able to determine the normal modes of a system from time-dependent numerical simulations is a practical effort. A spectral analysis of the variables at different points in the spatial domain do give a good indication of the frequencies present in the results, but the very relevant spatial structure of the associated eigenmodes cannot be easily achieved with such analysis. Hence, a means of extracting the spatial profile of eigenfunctions together with their associated oscillatory frequencies from time-dependent simulations is desirable. In this way, the results can be compared to observations to ascertain the presence of a given normal mode in the coronal structure under study. The Complex Empirical Orthogonal Function \citep[CEOF; see][]{vonstorch1999,hannachi2007} analysis is a tool that satisfies these requirements: it takes as its input the numerical values of one or more variables over a spatial domain and for a given time span, and returns the spatial and temporal information about the main modes of variability contained in the data, which in our case will be the main eigenmodes present in the time-dependent numerical simulations. Thus, the aim of retrieving the normal mode features is feasible with this procedure.

The main advance of this paper is the repeated application of the described combination of time-dependent numerical simulations and CEOF analysis. The later results allow to determine initial conditions (for the numerical simulations) that more accurately resemble those of the normal mode, resulting in a numerical simulation in which the amplitude of all other normal modes is reduced with respect to the previous iteration. Therefore, a repetition of this process leads to successively better approximations to a normal mode and convergence to a prescribed accuracy can be achieved. Since our aim is to test the feasibility of the new method, we keep our model as simple as possible, considering a slab loop model and neglecting the model improvements mentioned above (coronal loop curvature, longitudinal density stratification, magnetic field expansion, \ldots) In the presentation of the iterative method we follow a textbook approach: a simple test case with known solution is used, approximate solutions are found, the evolution of the error with the iterations is studied, and a proxy for this error that can be used in the stopping criterion is defined in terms of two successive approximations to the solution.

The outline of this paper is as follows: the equilibrium configuration and the equations for small amplitude perturbations are presented in Section~\ref{sec:equil_eqns}. Analytical expressions for the normal modes of this system are introduced in Section~\ref{sec:normal_modes}. The time-dependent equations are solved in Section~\ref{sec:simulation} for a prescribed initial condition and the CEOF analysis is applied to the results of this simulation; hence, the first iteration is complete, which allows us to give an approximation to the normal mode eigenfunctions and eigenfrequency. We next apply repeatedly the last two steps in an iterative process that improves the accuracy of the normal mode approximation (Section~\ref{sec:iterative}). Our conclusions are finally discussed in Section~\ref{sec:conclusions}.

\section{EQUILIBRIUM AND ZERO-$\beta$ GOVERNING EQUATIONS}
\label{sec:equil_eqns}

\begin{figure}[h]
  \centerline{
    \includegraphics[width=0.4\textwidth]{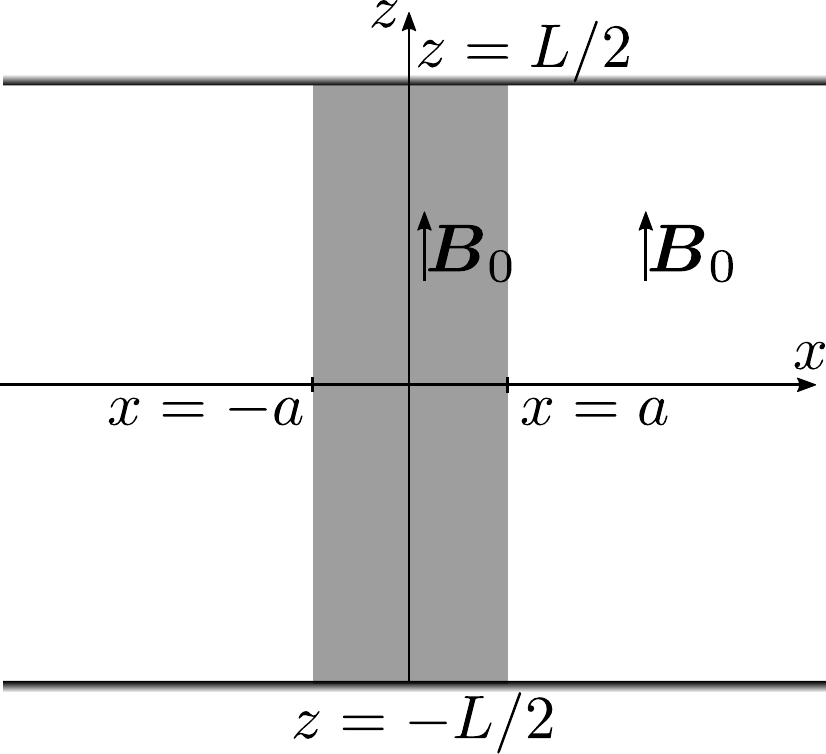}
  }
  \caption{Sketch of the equilibrium configuration, made of a plasma slab (hatched area) of width $2a$, length $L$, and density $\rhoi$ embedded in an environment with density $\rhoe$.}
  \label{fig:equil}
\end{figure}

We here use the Cartesian coordinate system shown in Figure~\ref{fig:equil}. The equilibrium is invariant in the $y$-direction and consists of a dense plasma slab of width $2a$ that extends between $x=-a$ and $x=a$ and is embedded in a rarer environment that fills the space $|x|>a$. The whole system is bounded by the two planes $z=\pm L/2$, with $L$ the slab length. In the equilibrium the magnetic field is uniform and points in the direction of the slab axis: $\vecBO=\BO\ez$; in addition, the plasma is at rest. This configuration has been often used to study the oscillations of a coronal loop. The $x$- and $z$-coordinates represent the directions transverse and longitudinal to the loop, respectively.

The equilibrium density is expressed as

\begin{equation}\label{eq:rho}
\rhoO(x) = 
   \begin{cases}
    \rhoi, & |x| \leq a, \\
    \rhoe, & |x| > a.
  \end{cases}
\end{equation}

\noindent The internal (i.e., inside the slab) and external Alfv\'en velocities are

\begin{equation}\label{eq:va}
\va(x) = 
   \begin{cases}
    \vai\equiv\frac{\BO}{\sqrt{\mu\rho_{\scriptstyle i}}}, & |x| \leq a, \\
    \vae\equiv\frac{\BO}{\sqrt{\mu\rho_{\scriptstyle e}}}, & |x| > a,
  \end{cases}
\end{equation}

\noindent with $\mu$ the permeability of free space.

We next introduce perturbations whose evolution is described by the ideal MHD equations, that in the zero-$\beta$ limit (i.e., zero plasma pressure) and in the absence of gravity read \citep{priest2014}

\begin{equation}\label{eq:dens_cont}
\frac{\partial\rho}{\partial t} = -\nabla\cdot(\rho\vecV),
\end{equation}

\begin{equation}\label{eq:mom_cont}
\rho\frac{\partial\vecV}{\partial t} = -\rho(\vecV\cdot\nabla)\vecV + \frac{1}{\mu}\left(\nabla\times\vecB\right)\times\vecB,
\end{equation}


\begin{equation}\label{eq:induction}
\frac{\partial\vecB}{\partial t} = \nabla\times\left(\vecV\times\vecB\right).
\end{equation}

\noindent Here $\rho$, $\vecV$, and $\vecB$ are the total (equilibrium plus perturbed) density, velocity, and magnetic field. Assuming small amplitude perturbations, Equations~(\ref{eq:dens_cont})--(\ref{eq:induction}) can be linearized. The density perturbation is only present in the first of these equations, so that it is a secondary quantity that can be obtained once the velocity ($\vecv$) and magnetic field ($\vecb$) perturbations are known. The linearized momentum and induction equations can be expressed as follows:

\begin{equation}\label{eq:linear_mom_cont}
\rhoO\frac{\partial\vecv}{\partial t} = \frac{1}{\mu}\left(\nabla\times\vecb\right)\times\vecBO,
\end{equation}

\begin{equation}\label{eq:linear_induction}
\frac{\partial\vecb}{\partial t} = \nabla\times\left(\vecv\times\vecBO\right),
\end{equation}

\noindent where $\vecv$ and $\vecb$ are both functions of position and time.

Now, perturbations are assumed to propagate in the $y$-direction with wavenumber $k_y$ and so the $y$-dependence of $\vecv$ and $\vecb$ is of the form $\exp(-ik_yy)$. The Cartesian components of Equations~(\ref{eq:linear_mom_cont}) and (\ref{eq:linear_induction}) then reduce to\footnote{The right-hand side of the $z$-component of Equation~(\ref{eq:linear_mom_cont}) is equal to zero and so it leads to $\vz=0$.}

\begin{equation}\label{eq:vx}
\frac{\partial\vx}{\partial t} = \frac{\BO}{\mu\rhoO}\left(\frac{\partial\bx}{\partial z}-\frac{\partial\bz}{\partial x}\right),
\end{equation}

\begin{equation}\label{eq:vy}
\frac{\partial\vy}{\partial t} = \frac{\BO}{\mu\rhoO}\left(\frac{\partial\by}{\partial z}+ik_y\bz\right),
\end{equation}

\begin{equation}\label{eq:bx}
\frac{\partial\bx}{\partial t} = \BO\frac{\partial\vx}{\partial z},
\end{equation}

\begin{equation}\label{eq:by}
\frac{\partial\by}{\partial t} = \BO\frac{\partial\vy}{\partial z},
\end{equation}

\begin{equation}\label{eq:bz}
\frac{\partial\bz}{\partial t} = -\BO\left(\frac{\partial\vx}{\partial x}-ik_y\vy\right).
\end{equation}

\noindent The velocity and magnetic field perturbations in these expressions are $\vecv(x,z,t)=\vx(x,z,t)\ex+\vy(x,z,t)\ey$ and $\vecb(x,z,t)=\bx(x,z,t)\ex+\by(x,z,t)\ey+\bz(x,z,t)\ez$.

In this paper we impose that the slab has a finite length, $L$, in the $z$-direction (Figure~\ref{fig:equil}) and that its ends are line-tied, that is, that the velocity perturbations are zero there. Moreover, in what follows we use the parameter values $\rhoi/\rhoe=10$, $L=50a$, and $k_ya=0.5$. Dimensionless values are obtained with the help of the length $a$, the velocity $\vai$, and the time $\tauai=a/\vai$.

\section{NORMAL MODES}
\label{sec:normal_modes}

Given that the plasma properties are uniform along the slab, the $z$-dependence of $\vx$ and $\vy$ is $\cos(k_zz)$. Equations~(\ref{eq:vx})--(\ref{eq:bz}) then reveal that the $z$-dependence of $\bz$ is $\cos(k_zz)$, while that of $\bx$ and $\by$ is $\sin(k_zz)$. To satisfy the boundary conditions at the slab ends, $k_z$ must be equal to $(n+1)\pi/L$, with $n=0$ for the longitudinally fundamental mode, $n=1$ for its first longitudinal overtone, etc. To study normal modes a temporal dependence of the form $\exp(i\omega t)$ is also imposed and so the perturbed velocity and magnetic field components are (the $y$-dependence is omitted)

\begin{align}
\label{eq:pert_normal_modes1}
\vx(x,z,t) &= \hvx(x)\cos(k_zz)e^{i\omega t}, \quad \vy(x,z,t) = \hvy(x)\cos(k_zz)e^{i\omega t}, \\
\label{eq:pert_normal_modes2}
\bx(x,z,t) &= \hbx(x)\sin(k_zz)e^{i\omega t}, \quad \by(x,z,t) = \hby(x)\sin(k_zz)e^{i\omega t}, \\
\label{eq:pert_normal_modes3}
\bz(x,z,t) &= \hbz(x)\cos(k_zz)e^{i\omega t}.
\end{align}

\noindent Equations~(\ref{eq:vx})--(\ref{eq:bz}) now reduce to

\begin{equation}\label{eq:vx_normal_mode}
\omega\hvx = -\frac{\BO}{\mu\rhoO}\left[k_z(i\hbx)-\frac{d}{d x}(i\hbz)\right],
\end{equation}

\begin{equation}\label{eq:vy_normal_mode}
\omega(i\hvy) = \frac{\BO}{\mu\rhoO}\left[k_z\hby+k_y(i\hbz)\right],
\end{equation}

\begin{equation}\label{eq:bx_normal_mode}
\omega(i\hbx) = -\BO k_z\hvx,
\end{equation}

\begin{equation}\label{eq:by_normal_mode}
\omega\hby = \BO k_z(i\hvy),
\end{equation}

\begin{equation}\label{eq:bz_normal_mode}
\omega(i\hbz) = -\BO\left[\frac{d\hvx}{d x}-k_y(i\hvy)\right].
\end{equation}

\noindent Now, the problem is to compute the $x$-dependence of the eigenfunctions $\hvx$, $i\hvy$, $i\hbx$, $\hby$, and $i\hbz$, which are all real, and the eigenvalue $\omega$.

It is straightforward to eliminate all variables in favor of $\hvx$, which leads to the following ordinary differential equation,

\begin{equation}\label{eq:ode_vx_normal_mode}
\frac{d^2\hvx}{d x^2} = m^2\hvx,
\end{equation}

\noindent with

\begin{equation}\label{eq:m}
m^2 = k_y^2+k_z^2-\frac{\omega^2}{\va^2}.
\end{equation}

\noindent The parameter $m$ takes the value $\mie$ when the Alfv\'en speed is substituted by its value $\vaie$ inside and outside the slab, respectively. After determining $\hvx$ one can obtain $i\hvy$ and $i\hbz$ from

\begin{equation}\label{eq:vy_from_vx_normal_mode}
i\hvy = \frac{k_y}{m^2}\frac{d\hvx}{d x},
\end{equation}

\begin{equation}\label{eq:bz_from_vx_normal_mode}
\frac{i\hbz}{\BO} = -\frac{1}{\omega}\frac{\kappa^2}{m^2}\frac{d\hvx}{d x},
\end{equation}

\noindent where

\begin{equation}\label{eq:kappa}
\kappa^2 = k_z^2-\frac{\omega^2}{\va^2} \equiv m^2-k_y^2.
\end{equation}

\noindent Again, $\kappa$ takes the values $\kappaie$ inside and outside the slab, respectively. The eigenfunctions $i\hbx$ and $\hby$ follow from Equations~(\ref{eq:bx_normal_mode}) and (\ref{eq:by_normal_mode}), and are just proportional to $\hvx$ and $i\hvy$, respectively.

To solve Equation~(\ref{eq:ode_vx_normal_mode}) one must impose boundary conditions in the $x$-direction, together with the proper jump conditions at the $x=\pm a$ interfaces, which according to, e.g., \citet{goedbloed2004} are the continuity of the normal velocity component ($\hvx$) and of the total pressure, which in turn leads to the continuity of $i\hbz$.

Because of the symmetry\footnote{The imposed boundary conditions in the $x$-direction are also symmetric: see Sections~\ref{sec:evanes_normal_modes} and \ref{sec:confined_normal_modes}.} of the equilibrium and of Equations~(\ref{eq:vx_normal_mode})--(\ref{eq:bz_normal_mode}) with respect to $x=0$, eigenfunctions are either even or odd: for kink modes $\hvx$ and $i\hbx$ are even about the slab axis, while $i\hvy$, $\hby$, and $i\hbz$ are odd; for sausage modes, the parity of the 5 eigenfunctions is the opposite. In our simulations only kink solutions are excited and so we restrict our analysis to these normal modes.

\subsection{Laterally evanescent normal modes}
\label{sec:evanes_normal_modes}

\citet{arregui2007} solved the eigenproblem of Equations~(\ref{eq:vx_normal_mode})--(\ref{eq:bz_normal_mode}) for solutions that are laterally evanescent, that is, for which the perturbations vanish as $x\rightarrow\pm\infty$; see their Section~3 and also \citet{roberts1981} for the treatment of the $k_y=0$ case. The kink solution that satisfies these constraints has the following $x$-velocity component:

\begin{equation}\label{eq:vx_evanescent}
\hvx(x) =
  \begin{cases}
    C\exp(\me x), & \text{for } x<-a, \\
    A\cosh(\mi x), & \text{for } -a\leq x\leq a, \\
    C\exp(-\me x), & \text{for } x>a,
  \end{cases}
\end{equation}

\noindent where the positive value of $\me$ is taken and


\begin{equation}\label{eq:C_evanescent}
C = A \exp(\me a)\cosh(\mi a).
\end{equation}

\noindent The constant $A$ can be arbitrarily chosen and so we set $A=1$. The eigenfrequency is the solution to the dispersion relation

\begin{equation}\label{eq:dr_evanescent}
\tanh(\mi a) = -\frac{\kappae^2}{\kappai^2}\frac{\mi}{\me}.
\end{equation}

\begin{figure}[ht!]
  \centering{\includegraphics[width=0.7\textwidth,angle=0]{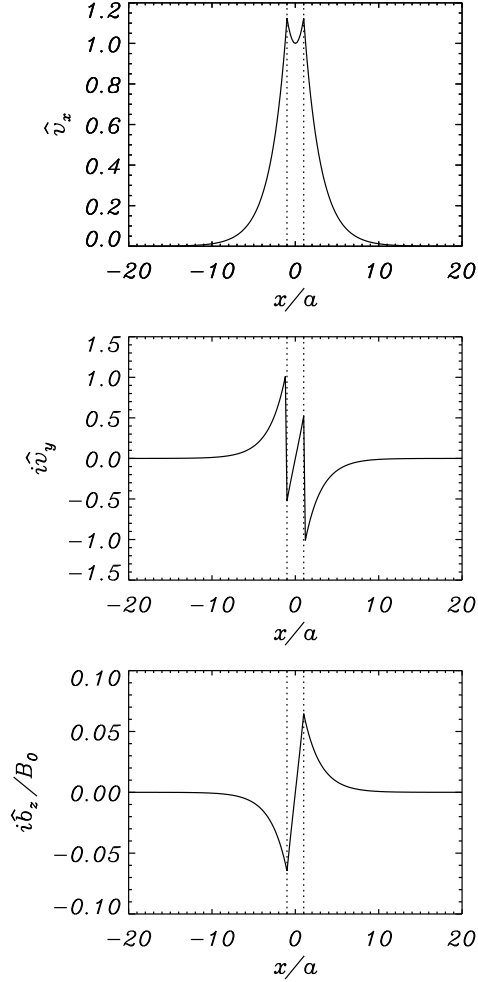}}
  \caption{Normal mode: from top to bottom, eigenfunctions $\hvx$, $i\hvy$, and $i\hbz$ for the fundamental evanescent kink mode. The two dotted lines correspond to the slab boundaries.}
  \label{fig:kink_eigenfunctions}
\end{figure}

Figure~\ref{fig:kink_eigenfunctions} displays the eigenfunctions $\hvx$, $i\hvy$, and $i\hbz$ for the fundamental kink mode. They possess the parity and continuity properties described above: $\hvx$ is even, $i\hvy$ and $i\hbz$ are odd, and $\hvx$ and $i\hbz$ are continuous at the interfaces $x=\pm a$. On the other hand, $i\hvy$ jumps abruptly at these boundaries. In addition, these functions decay exponentially with $x$, as described by Equation~(\ref{eq:vx_evanescent}). The longitudinal harmonics have a similar spatial structure of eigenfunctions. The frequencies for the $n=0,2,4$ longitudinal harmonics are $\omega/\tauai=0.1011,0.2989,0.4852$.

\subsection{Laterally confined normal modes}
\label{sec:confined_normal_modes}

In Section~\ref{sec:simulation} we solve numerically the initial value problem made of Equations~(\ref{eq:vx})--(\ref{eq:bz}) with suitable initial and boundary conditions. We consider the spatial domain $-20a\leq x\leq 20a$. Given that the boundaries are sufficiently far from the slab, the evanescent eigensolution of Figure~\ref{fig:kink_eigenfunctions} is almost zero at the edges of the numerical domain and so it is, in practice, a solution to the initial value problem. Placing the boundaries at a finite distance from the slab, however, adds new, non-evanescent eigensolutions that can be excited by the initial perturbation. By replicating the analysis of Section~\ref{sec:evanes_normal_modes} with the boundary condition $\hvx=0$ at $x=\pm\Lx$ the laterally confined eigenfunctions can be obtained. It will suffice to say that the fundamental confined mode has $\omega=1.676/\tauai$.

\section{TIME-DEPENDENT NUMERICAL SIMULATIONS}
\label{sec:simulation}

\subsection{Simulation setup and numerical method}
\label{sec:simulation_setup}

We now solve numerically Equations~(\ref{eq:vx})--(\ref{eq:bz}) in the region $-\Lx\leq x\leq\Lx$, $-L/2\leq z\leq L/2$; see Figure~\ref{fig:equil}. The coefficients of the system of partial differential equations can be made real by using the independent variables $\vx$, $i\vy$, $\bx$, $i\by$, and $\bz$. Our initial disturbance is such that the full slab is subject to an initial transverse forcing given by

\begin{equation}\label{eq:initial_disturbance}
\vx(x,z,t=0)=\vO\exp\left(-\frac{x^2}{a^2}\right)\exp\left(-\frac{z^2}{a^2}\right),
\end{equation}

\noindent while all other variables are initially zero. This initial perturbation represents a sudden deposition of energy at the slab center. The $v_x$ perturbation is even about $x=0$ and so can only excite kink modes. Since the transverse profile of $\vx$ resembles that of the laterally evanescent kink modes (top panel of Figure~\ref{fig:kink_eigenfunctions}), a large portion of the energy in the initial disturbance will go to these modes. But one can expect that the laterally confined mode of Section~\ref{sec:confined_normal_modes} will also be excited. 

The simulation box is determined by the lengths $\Lx=20a$ and $L=50a$, and a uniform grid of 4001$\times$51 points in the $x$- and $z$-directions is used. The grid is coarser along the slab because it is sufficient to capture well the smooth sinusoidal dependence of normal modes in the $z$-direction; on the other hand, the grid is much finer across the slab because normal modes have much more structure in this direction. The numerical simulation is stopped at $t\simeq 280\tauai$, which is $\sim 4.5$ and $\sim 75$ periods of the fundamental laterally evanescent and confined modes, respectively. The time step is $\Delta t=0.704\tauai$.

The numerical method used to solve the linearized wave equations is based on the method of lines (MOL). Time and space are treated independently, using a third order Runge-Kutta method and a six order finite difference method, respectively. Artificial dissipation is included to avoid oscillations on the grid scale. This method has been used successfully in the past \citep[e.g.,][]{bona2009} and has a weak effect on the attenuation of the physical oscillations reported in the simulations. Since the linear hyperbolic MHD equations are solved explicitly, the time step is subject to the CFL condition. Note that in the linearized MHD equations there are terms proportional to $k_y$: these terms are incorporated to the code as simple source terms.

Although we solve the linearized MHD equations, there are jumps in the perturbed variables (in $i v_y$ and $b_y$) due to the discontinuities in the equilibrium variables. We have decided to use a simple numerical scheme that is not shock-capturing (better suited for discontinuities) since the effect of the jump is rather small in the temporal evolution of the different quantities.

Line-tying conditions are applied at $z=\pm L/2$, meaning that the velocities are zero, while for the rest of variables the derivatives with respect to $z$ are zero. At $x=\pm L_x$ we impose that the derivatives with respect to $x$ of all the variables are zero. This condition does not allow a perfect outward transmission of the waves and some reflections are produced. A direct consequence of these reflections is the presence of the laterally confined normal mode in our simulations.

\subsection{Results}
\label{sec:simulation_results}

\begin{figure}[ht!]
  \centering{\includegraphics[width=0.7\textwidth,angle=0]{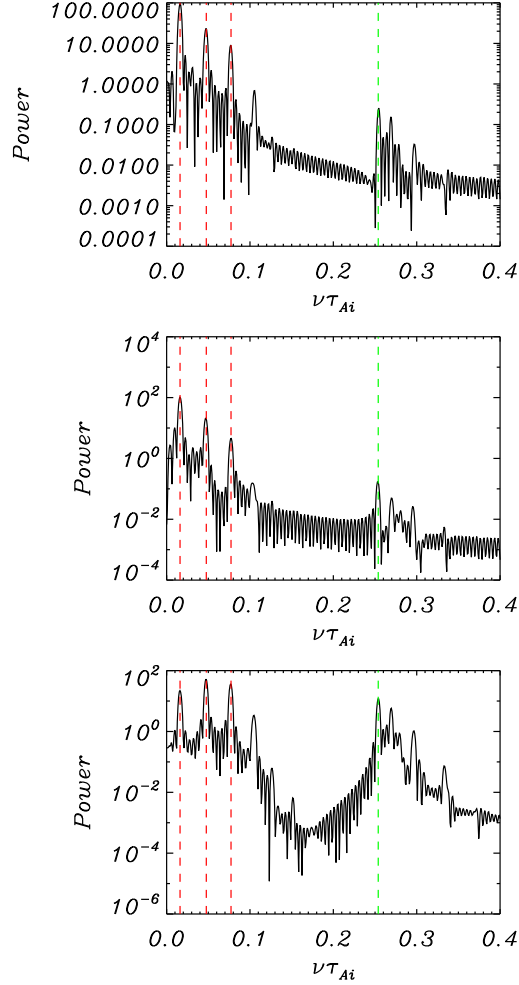}}
  \caption{Numerical simulation: Lomb-Scargle periodogram of $\vx$ at position $x=0$, $z=0$ (top),  $i\vy$ at position $x=a$, $z=0$ (middle), and  $\bz$ at position $x=a$, $z=0$ (bottom). To compute the power spectra only data for $t\geq50\tauai$ are kept so as to remove the effect of the transient in the frequency estimation. Vertical red (green) lines are drawn at the frequencies of the first laterally evanescent (laterally confined) harmonics.}
  \label{fig:periodogram}
\end{figure}

The initial condition excites a large number of longitudinal harmonics, both evanescent and confined in the $x$-direction, together with leaky waves that travel away from the slab. The emission of these leaky waves is clear until $t\simeq 50\tauai$, after which only the kink normal modes of Sections~\ref{sec:evanes_normal_modes} and \ref{sec:confined_normal_modes} remain. Evidence of the presence of these normal modes comes from the spectral analysis of $\vx$, $i\vy$, and $\bz$ at a given location, which is selected so that normal modes have a non-negligible amplitude. Thus, for the transverse velocity component, $\vx$, we choose the point $x=0$, $z=0$, while for the other two variables the position $x=a$, $z=0$ is preferred. The Lomb-Scargle periodograms \citep{lomb1976,scargle1982} at these points are shown in Figure~\ref{fig:periodogram}. The three panels display the largest power peaks at $\nu=0.01601/\tauai$ (i.e., $\omega=0.1006/\tauai$), $\nu=0.04721/\tauai$ ($\omega=0.2966/\tauai$), and $\nu=0.07681/\tauai$ ($\omega=0.4826/\tauai$); these values are in excellent agreement with those of the lowest three laterally evanescent harmonics. The periodograms also show large power above $\nu=0.2/\tauai$ caused by the excitation of laterally confined normal modes. Indeed, the largest peak in this frequency range is at $\nu=0.2544/\tauai$ (i.e., $\omega=1.598/\tauai$), again in good agreement with the value quoted in Section~\ref{sec:confined_normal_modes}. It is worth noting that the power at $\omega=0.1011/\tauai$ is 2--3 orders of magnitude higher than that at $\omega=1.676/\tauai$ for $\vx$ and $i\vy$, although in the case of $\bz$ the two peaks are of similar magnitude. The reason for this is that the height of a power peak comes from the combination of the energy deposited by the initial disturbance in each normal mode (which is much larger for the evanescent one) and the amplitude of each variable (which in the case of $\bz$ is much smaller for the evanescent normal mode than for the confined one). The combination of these two factors results in the function $\bz$ containing similar power in the evanescent and confined normal modes in this numerical simulation.

\section{COMPLEX EMPIRICAL ORTHOGONAL FUNCTION ANALYSIS}
\label{sec:ceof}

In this section we go beyond the normal mode frequency we just obtained and attempt to determine the normal mode eigenfunction structure using the Complex Empirical Orthogonal Function (CEOF) analysis. We will here give a very brief summary of this method; a detailed description can be found in \citet{horel1984,wallace1972,vonstorch1999}; \citet[][where it is called Complex Hilbert EOF]{hannachi2007} and an application to the study of coronal oscillations in \citet{terradas2004}.

The CEOF analysis is a numerical method that takes as its input a field $U(\vec{r},t_l)$ discretized over a spatial mesh of points $\vec{r}=(x_i,y_j,z_k)$ and evaluated at the discrete times $t_l$. Its output is a set of CEOF modes, which are not necessarily associated to physical modes of the system under study, each of them described by four measures called the temporal amplitude and phase and the spatial amplitude and phase. Together with these measures, the CEOF analysis associates to each mode a fraction of the total field variance. Once the CEOF code is fed with the input field, the ``highest contributing'' CEOF mode, that is, the one associated to the largest fraction of the total field variance, is retrieved first and other CEOF modes are obtained next in decreasing order of their fraction of the total field variance. The sum of the modes' fraction of the total field variance tends to 1 as the number of CEOF modes is increased. The execution is stopped when the percentage of the total field variance accounted for by all the retrieved CEOF modes exceeds a pre-established value, here taken as 99.9\%.

In our case the field $U$ can be, for example, the velocity component $\vx(x_i,z_k,t_l)$ obtained in the numerical simulation of Section~\ref{sec:simulation}. This means that the input field is a three-dimensional data cube. Such as mentioned above, each of the obtained CEOF modes has empirically computed temporal and spatial measures, called the temporal amplitude, $R(t_l)$, the temporal phase, $\phi(t_l)$, the spatial amplitude, $S(x_i,z_k)$, and the spatial phase, $\theta(x_i,z_k)$. The spatial and temporal variability of the field described by this CEOF mode is

\begin{equation}\label{eq:ceof_mode}
Re\left\{R(t_l)\exp[i\phi(t_l)]S(x_i,z_k)\exp[-i\theta(x_i,z_k)]\right\},
\end{equation}

\noindent where $Re$ denotes the real part. A CEOF mode that, for example, represents a propagating wave has a temporal phase that varies linearly with $t_l$ and a spatial phase that varies linearly with $x_i$ and $z_k$. The CEOF modes that we expect to find when analyzing the results of the numerical simulation, however, are standing waves. In this case, the temporal phase also varies linearly with $t_l$, but the spatial phase is such that two regions in which the difference of $\theta(x_i,z_k)$ is an integer multiple of $2\pi$ correspond to in-phase oscillations, while oscillations that are in anti-phase display a phase difference that is an odd multiple of $\pi$. In our results we will also find that a standing wave can have a phase that slowly varies in space, which is nothing but a modulation of $S(x_i,z_k)$ by the factor $\exp[-i\theta(x_i,z_k)]$. Section~3 of \citet{terradas2004} gives simple two-dimensional examples of the outcome of the CEOF analysis when applied to a synthetic signal made of the sum of a propagating and a standing wave.

Our hypothesis is that the CEOF analysis applied to the results of the numerical simulation of Section~\ref{sec:simulation} will provide an approximation, by means of Equation~(\ref{eq:ceof_mode}), to the evanescent normal mode eigenfunctions. Given that the eigenfunctions do not depend on time, we will ignore the temporal variation given by the measures $R(t)$ and $\phi(t)$ in Equation~(\ref{eq:ceof_mode}) and will only retain the real part of the spatial measures. Let us take, for example, the variable $\vx$, which for a normal mode has the eigenfunction $\hvx(x)\cos(k_zz)$. The CEOF approximation to this eigenfunction is

\begin{equation}\label{eq:ceof_mode_compare_vx}
\vxceof(x_i,z_k) = \Svx(x_i,z_k)\cos\thetavx(x_i,z_k),
\end{equation}

\noindent where $\Svx$ and $\thetavx$ are the spatial amplitude and phase of $\vx$ and the tilde in $\vxceof$ indicates that this is an approximation to the normal mode $\vx$. A numerical comparison between the normal mode eigenfunction and its approximation from the CEOF analysis is obtained with the help of the difference

\begin{equation}\label{eq:ceof_error_vx}
\Dvx(x_i,z_k) = \hvx(x_i)\cos k_zz_k-\vxceof(x_i,z_k).
\end{equation}

\noindent An analogous expression can be built for all other eigenfunctions.

Regarding $i\vy$, its eigenfunction for the confined mode is $i\hvy(x)\cos(k_zz)$ with $i\hvy(x)$ given by Equations~(\ref{eq:vy_from_vx_normal_mode}) and (\ref{eq:vx_evanescent}). The CEOF approximation to this eigenfunction is

\begin{equation}\label{eq:ceof_mode_compare_vy}
i\vyceof(x_i,z_k) = \Sivy(x_i,z_k)\cos\thetaivy(x_i,z_k),
\end{equation}

\noindent with $\Sivy$ and $\thetaivy$ the spatial amplitude and phase of $i\vy$. The case of $\bz$ requires special attention. Its eigenfunction is $i\hbz(x)\cos(k_zz)$, where $i\hbz(x)$ can be obtained from Equations~(\ref{eq:bz_from_vx_normal_mode}) and (\ref{eq:vx_evanescent}). In the numerical simulation, however, we have not used the variable $i\bz$ but $\bz$. For this reason, the CEOF approximation to $i\bz$ requires inserting a factor $i$ inside $Re\left\{\ldots\right\}$ of Equation~(\ref{eq:ceof_mode}). We then have that the CEOF approximation to $i\bz$ is

\begin{equation}\label{eq:ceof_mode_compare_bz}
i\bzceof(x_i,z_k) = \Sbz(x_i,z_k)\sin\thetabz(x_i,z_k),
\end{equation}

\noindent where $\Sbz$ and $\thetabz$ are the spatial amplitude and phase of $\bz$. Concerning $\bx$ and $\by$, Equations~(\ref{eq:bx_normal_mode}) and (\ref{eq:by_normal_mode}) tell us that their respective CEOF approximations can be obtained from those of $\vx$ and $i\vy$. Finally, approximations to $\hvx(x)$, $i\hvy(x)$, and $i\hbz(x)$ can be derived by taking a cut along $z=z_k$ of Equations~(\ref{eq:ceof_mode_compare_vx}), (\ref{eq:ceof_mode_compare_vy}), and (\ref{eq:ceof_mode_compare_bz}).

Before applying the CEOF method to the results of our simulation, two more comments are in order. First, the transient phase is excluded from the analysis by considering $t\geq50\tauai$ only. Second, to reduce the memory requirements and speed up the computation of the CEOF modes, the values of $\vx$, $i\vy$ and $\bz$ are interpolated from the 4001$\times$51 numerical grid to a grid of $\Nx\times\Nz$ points (here we use $\Nx=201$, $\Nz=25$). To do so, in the $x$- and $z$-directions only 1 every 20 points and 1 every 2 points, respectively, from the numerical simulation are kept for the CEOF analysis.

\subsection{Results}
\label{sec:ceof_vxvybz}

The CEOF method has the possibility of analyzing several fields simultaneously, which allows a better characterization of the physical modes because more restrictions are imposed by the higher complexity of the combined fields. Thus, we run the CEOF code on the fields $\vx(x_i,z_k,t_l)$, $i\vy(x_i,z_k,t_l)$, and $\bz(x_i,z_k,t_l)$ together. To do this, the three data cubes are put next to each other and a larger data cube is created. We choose to join the three 201$\times$25$\times$330 cubes by attaching their $xt$-faces, so that the CEOF input is a cube of 201$\times$75$\times$330 data values. After the CEOF analysis is complete, we obtain a collection of CEOF modes, each of them characterized by its temporal amplitude and phase, $R(t)$ and $\phi(t)$, and its spatial amplitude and phase, $S(x,z)$ and $\theta(x,z)$, that can be split into the spatial measures $\Svx$ and $\thetavx$ of the field $\vx(x_i,z_k,t_l)$, $\Sivy$ and $\thetaivy$ of the field $i\vy(x_i,z_k,t_l)$, and $\Sbz$ and $\thetabz$ of the field $\bz(x_i,z_k,t_l)$. These measures can in turn be inserted into Equations~(\ref{eq:ceof_mode_compare_vx}), (\ref{eq:ceof_mode_compare_vy}), (\ref{eq:ceof_mode_compare_bz}) to obtain the approximate CEOF eigenfunctions.

\begin{figure}[ht!]
  \centering{\includegraphics[width=\textwidth,angle=0]{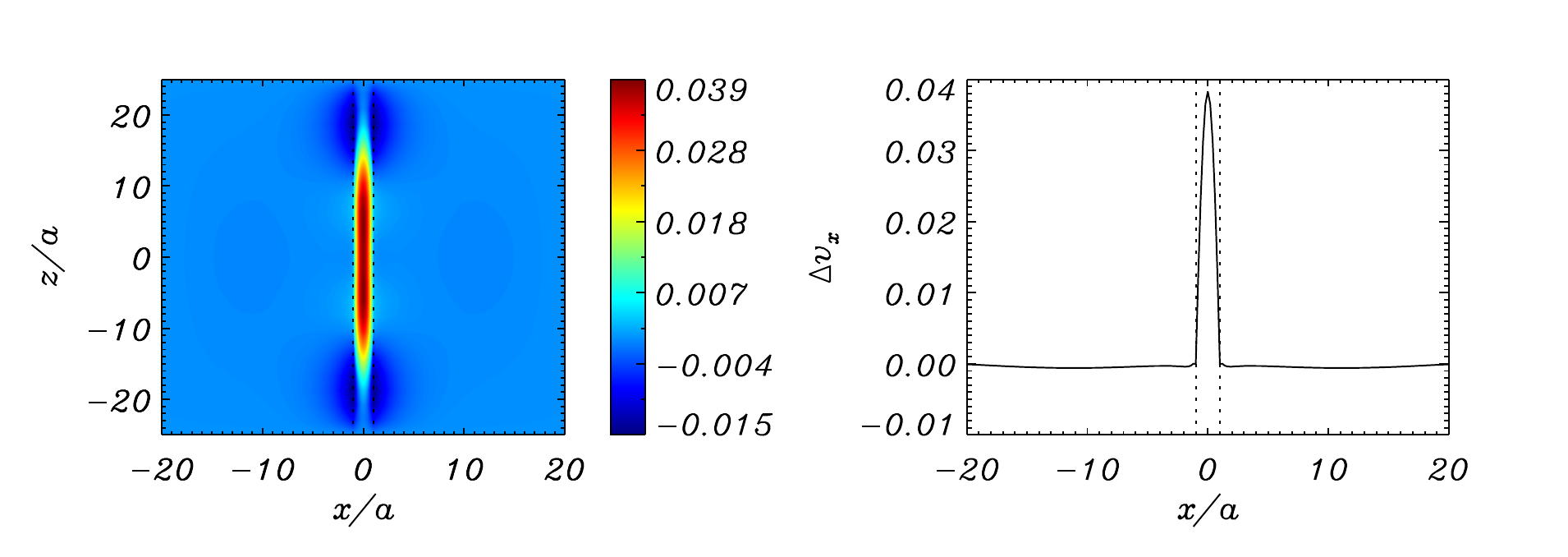}}
  \centering{\includegraphics[width=\textwidth,angle=0]{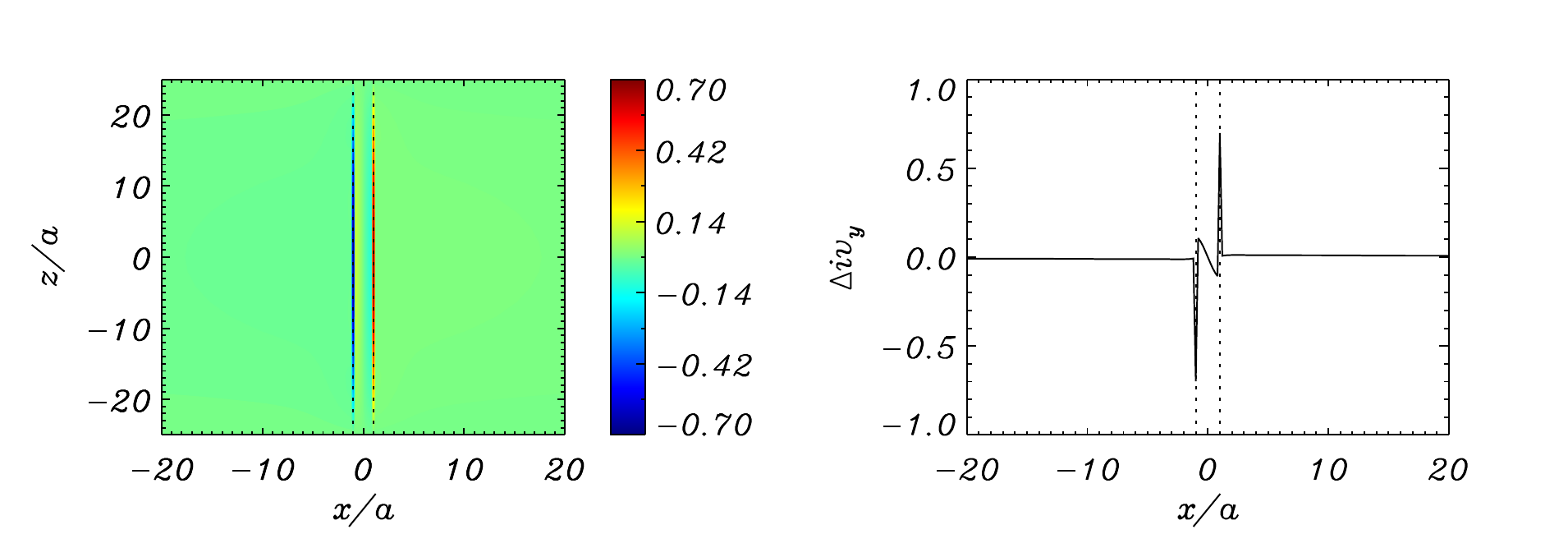}}
  \centering{\includegraphics[width=\textwidth,angle=0]{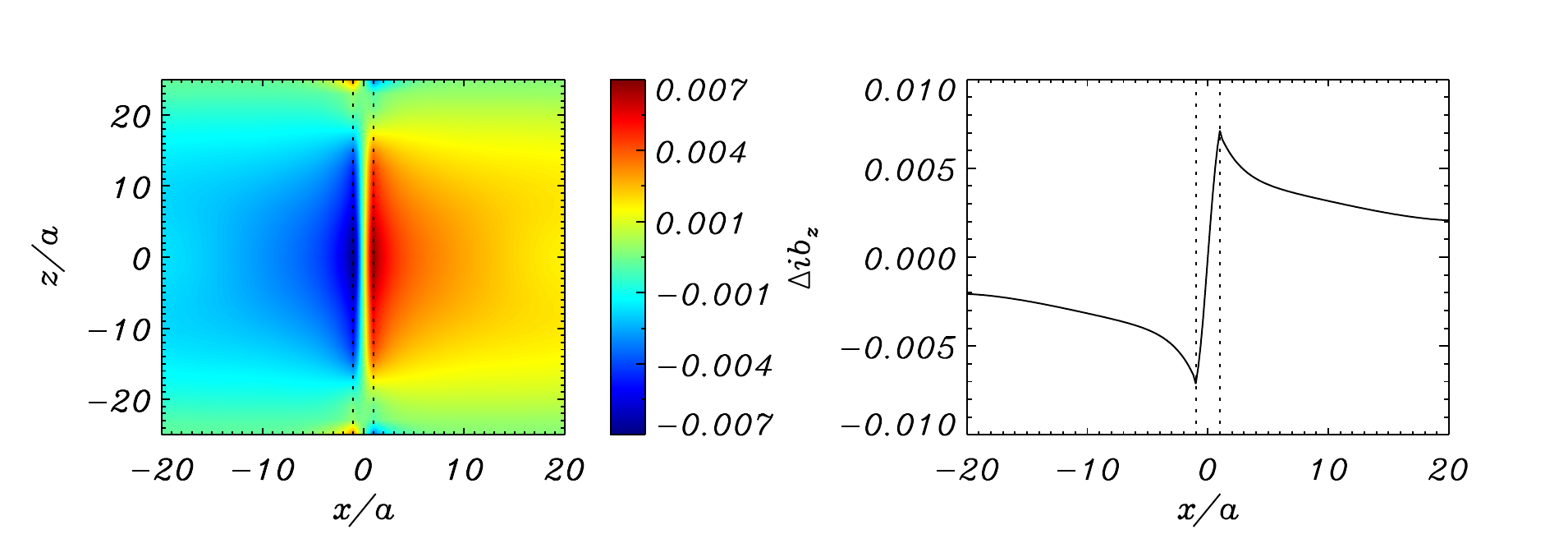}}
  \caption{CEOF analysis: difference $\Dvx$ (top row), $\Dvy$ (middle row), and  $\Dbz$ (bottom row) between the fundamental evanescent eigenfunctions and their approximation from the first CEOF mode. Left: two dimensional distribution of the difference. Right: cut of the difference along $z=0$. Dotted lines are plotted at the slab boundaries.}
  \label{fig:ceof_vxvybz_comparison}
\end{figure}

The first CEOF mode accounts for 64.8\% of the total field variance and corresponds to the fundamental evanescent normal mode. Its frequency is determined by fitting the straight line $\phi(t)=\omega t+\phi_0$ to the temporal phase, which yields $\omega=0.1016/\tauai$. This value is in excellent agreement with the normal mode frequency $\omega=0.1011/\tauai$. The goodness of the CEOF approximation to the normal mode can also be judged with the help of the differences $\Dvx$, $\Dvy$, $\Dbz$, which are presented in Figure~\ref{fig:ceof_vxvybz_comparison}. To make this figure, the $\hvx$ eigenfunction is normalized to a maximum value of 1 and the CEOF approximation $\vxceof$ is also normalized to 1 at the position where the eigenfunction is maximum. The conclusion from this figure is that the CEOF analysis of the numerical simulation results allows us to recover the normal mode eigenfunction $\hvx$ with an error below~4\%. We next turn our attention to the error of $i\vy$ and $i\bz$. The middle row of Figure~\ref{fig:ceof_vxvybz_comparison} gives the difference $\Dvy$. Except for the points on the boundaries, $x=\pm a$, the error is smaller than 10\% inside the slab ($|x|<a$) and practically zero outside the slab ($|x|>a$). The bottom row of Figure~\ref{fig:ceof_vxvybz_comparison} gives the difference $\Dbz$, which also attains its largest value, of the order of 15\% the eigenfunction value, at the slab boundary.

\begin{figure}[ht!]
  \centering{\includegraphics[width=\textwidth,angle=0]{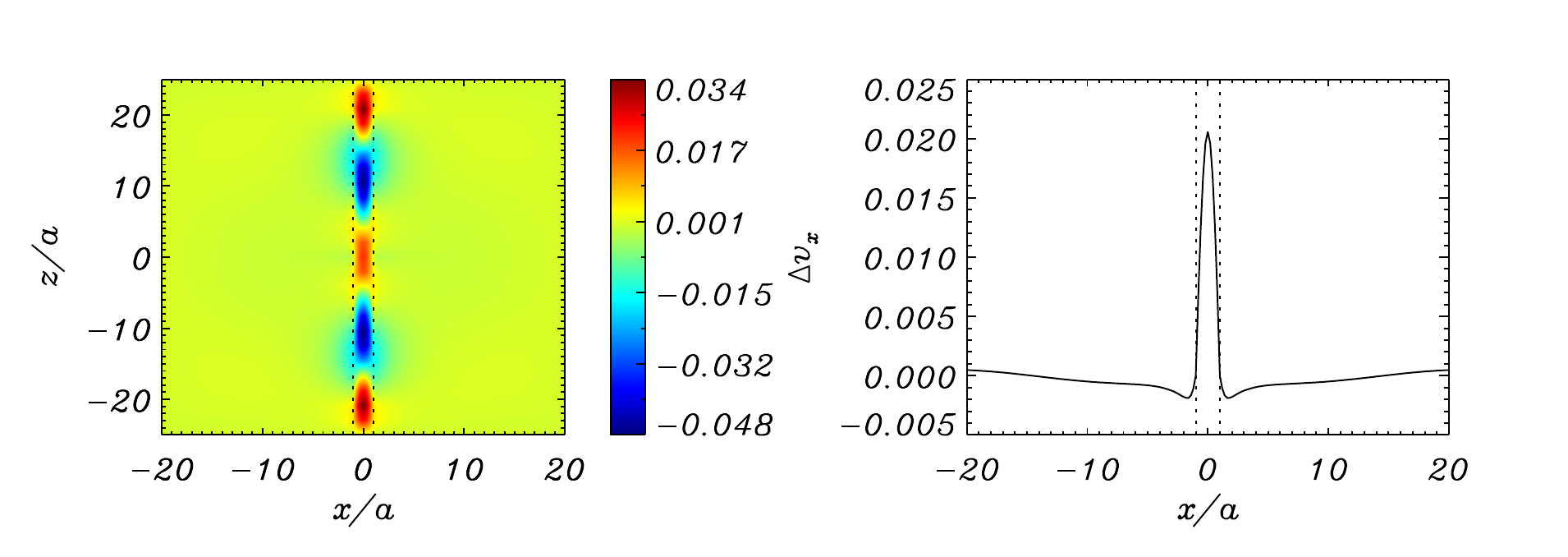}}
  \centering{\includegraphics[width=\textwidth,angle=0]{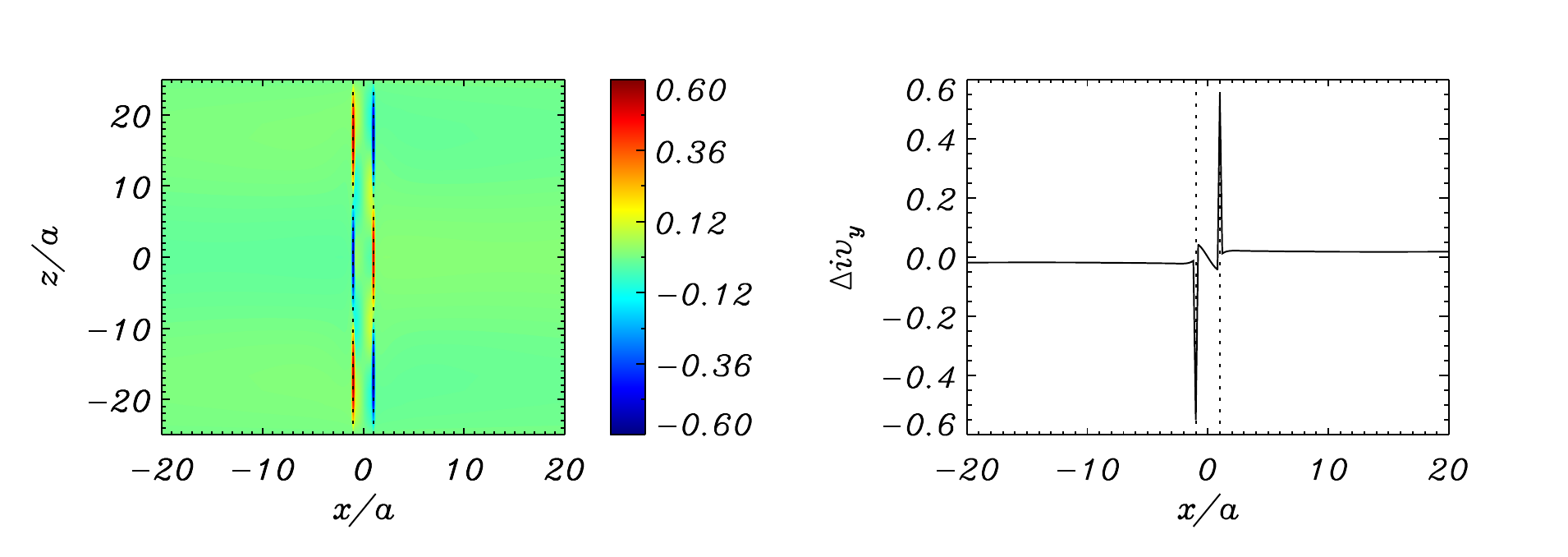}}
  \centering{\includegraphics[width=\textwidth,angle=0]{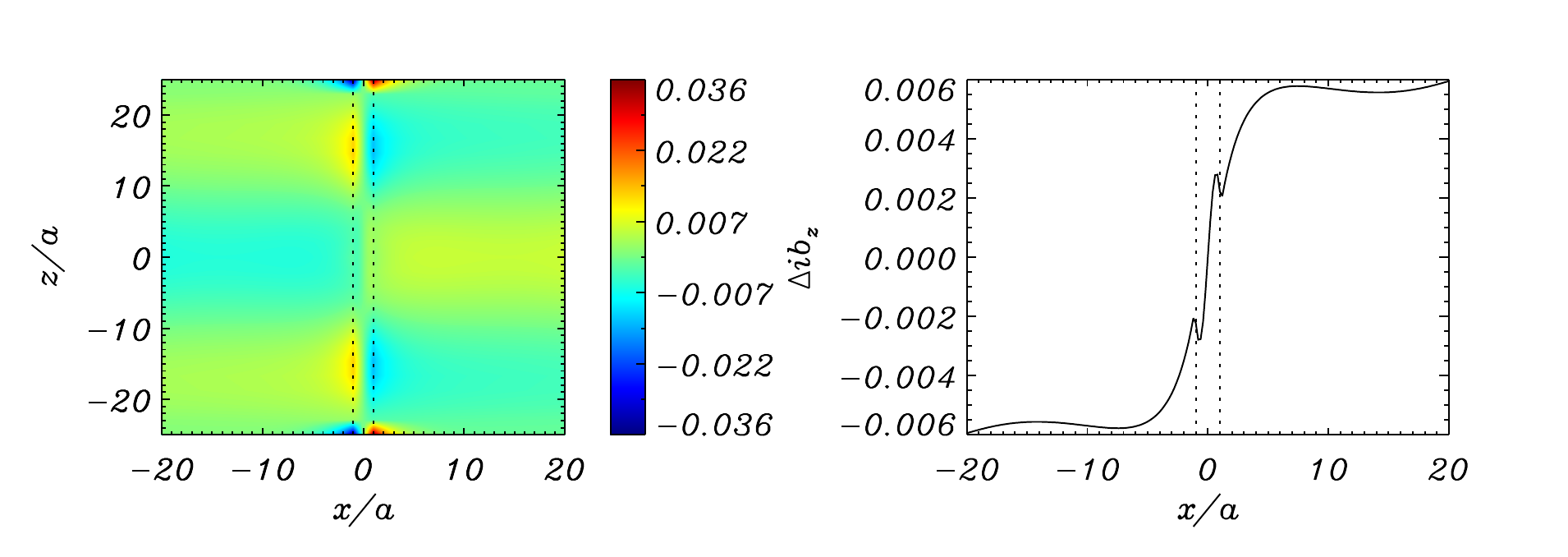}}
  \caption{CEOF analysis:  same as Figure~\ref{fig:ceof_vxvybz_comparison} for the second longitudinal evanescent overtone.}
  \label{fig:ceof_vxvybz_comparison_2nd}
\end{figure}

The second CEOF mode accounts for 15.9\% of the total field variance and corresponds to the second longitudinal evanescent overtone. A linear fit to the temporal phase results in the frequency $\omega=0.2969/\tauai$, which is very close to the analytical value $\omega=0.2989/\tauai$. Figure~\ref{fig:ceof_vxvybz_comparison_2nd} shows the difference between the normal mode and CEOF approximation to the eigenfunctions. We see that the error in the second longitudinal overtone is almost a factor of 2 better than that of the fundamental mode. When comparing the exactness of the CEOF approximation for the fundamental and the second overtone, we find a better agreement in the second case because there are more periods of this normal mode in the numerical simulation.

The conclusion of this section is that, while the CEOF approximation to $\hvx$ is acceptable, those to $i\hvy$ and $i\hbz$ are not too good. This situation will be improved by the application of the iterative method presented in the following section.



\section{ITERATIVE METHOD}
\label{sec:iterative}

\subsection{Description of the method}
\label{sec:iterative_description}

The scheme we have used so far consists of two steps: (i) a time-dependent numerical simulation of Equations~(\ref{eq:vx})--(\ref{eq:bz}) followed by (ii) the CEOF analysis of the obtained results. If we imagine that an eigenmode is perfectly described by a CEOF mode, then one could run a numerical simulation with initial conditions given by the eigenfunctions and so the obtained temporal evolution would be that of the eigenmode. At this point, this is not the case, but we have seen that the CEOF analysis produces an approximation to a normal mode eigenfunctions. We thus devise an iterative method that is made of the repeated application of steps (i) and (ii), in which the initial conditions of the numerical simulation of a given iteration are taken from the CEOF method of the previous iteration. The iterations will be stopped once a given measure of goodness is reached. The iterative method is carried out separately for each normal mode.

We first need to determine which information is required from the CEOF analysis to fix the initial conditions. Rather than using the time dependence $\exp(i\omega t)$ of Equations~(\ref{eq:pert_normal_modes1})--(\ref{eq:pert_normal_modes3}) we assume that $\vx(x,z,t)$ is maximum at $t=0$ and so it has the form $\vx(x,z,t) = \hvx(x)\cos(k_zz)\cos(\omega t)$. Now, Equations~(\ref{eq:vx})--(\ref{eq:bz}) tell us that $\vx$ and $i\vy$ are in phase (in time) and that they are a quarter of a period out of phase with respect to $\bx$, $i\by$, and $\bz$. This implies that $i\vy(x,z,t)$ is also maximum at $t=0$ and that the perturbed magnetic field components vanish at the start of the numerical simulation. Hence, the information that the CEOF analysis must provide to repeat step (i) is the approximation to $\hvx$ and $i\hvy$ provided by Equations~(\ref{eq:ceof_mode_compare_vx}) and (\ref{eq:ceof_mode_compare_vy}).

\subsection{Results}
\label{sec:iterative_results}

\begin{figure}[ht!]
  \centering{\includegraphics[width=0.49\textwidth,angle=0,trim={10ex 0 20ex 0},clip]{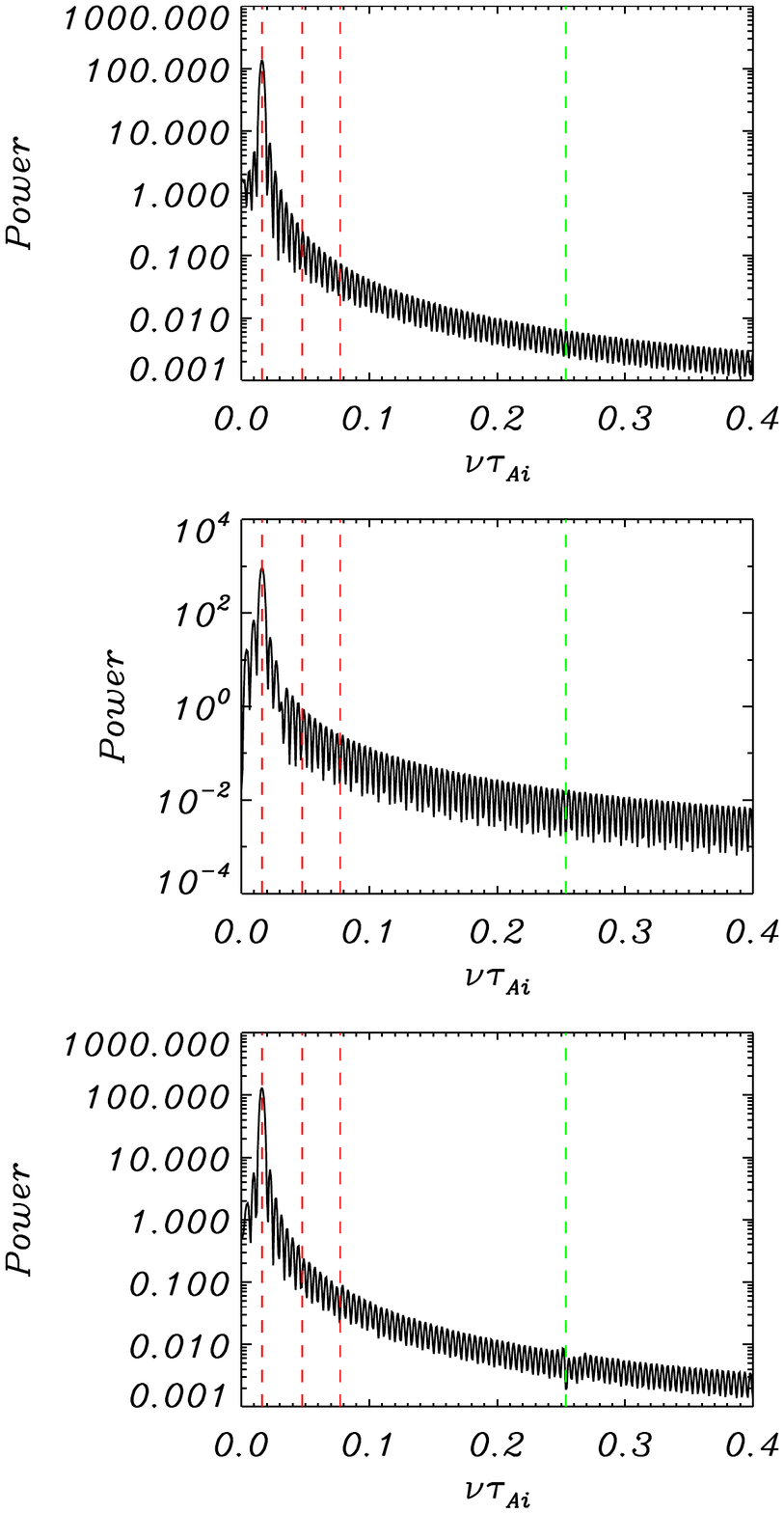} \includegraphics[width=0.49\textwidth,angle=0,trim={10ex 0 20ex 0},clip]{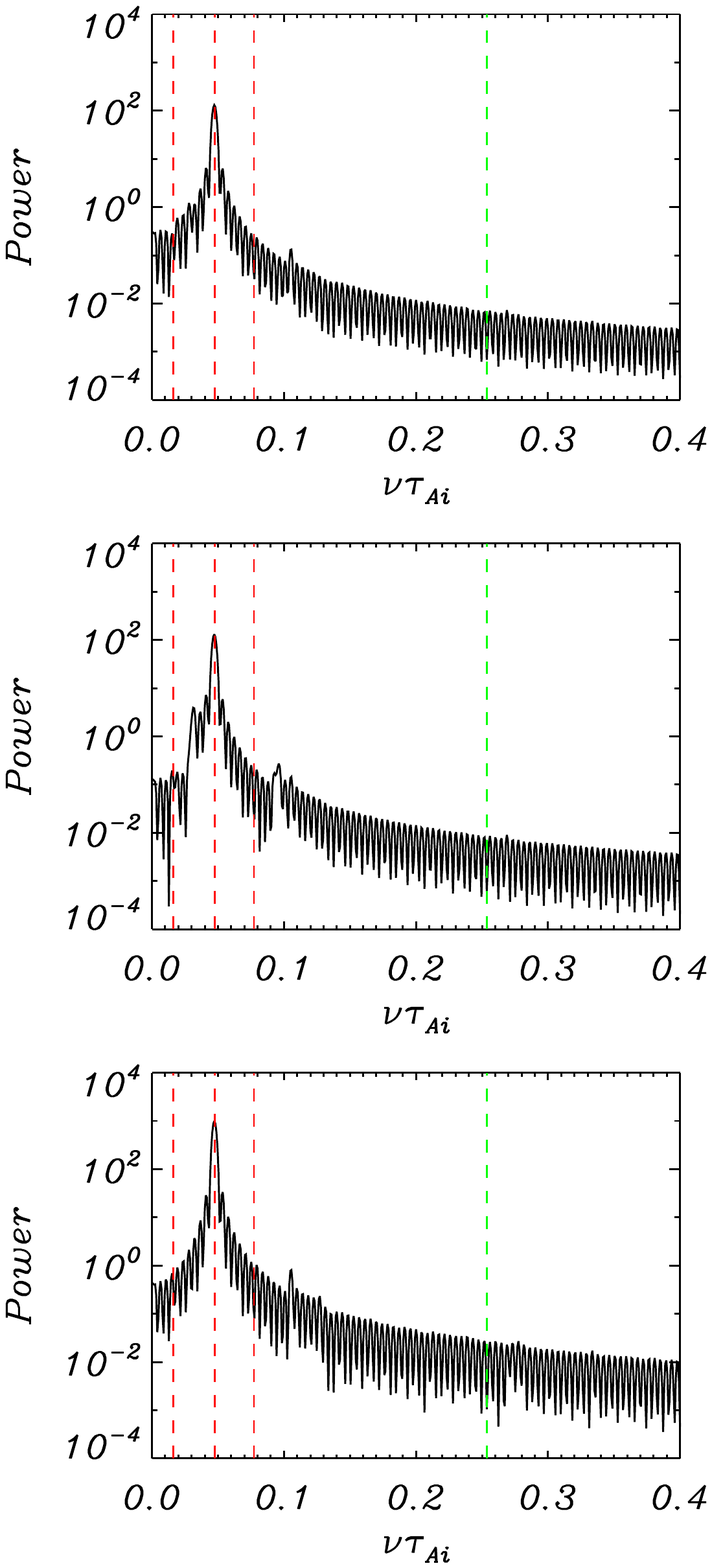}}
  \caption{Iterative method: same as Figure~\ref{fig:periodogram} for the numerical simulations of iteration \#6. Left:~fundamental evanescent mode, right: second longitudinal evanescent overtone.}
  \label{fig:periodogram_iter4}
\end{figure}

We are then ready to carry out the iterative process. Iteration \#1 consists of the numerical simulation of Section~\ref{sec:simulation} and the CEOF analysis of Section~\ref{sec:ceof_vxvybz}. The results we present now are a summary of the performance of 6 iterations, which are carried out independently for the fundamental evanescent mode and its second longitudinal overtone.
%
%
An excellent assessment of the performance of the iterative method can be gained from the power spectra of $\vx$, $i\vy$, and $\bz$ for the numerical simulation of the last iteration. These power spectra are shown in Figure~\ref{fig:periodogram_iter4} for the two normal modes. Whereas the numerical simulation of iteration \#1 displays power peaks at the frequencies of many harmonics, both simulations of iteration \#6 only show a power peak for a single normal mode. The left (right) panels of Figure~\ref{fig:periodogram_iter4} have non-negligible power around the maximum $\nu=0.01601/\tauai$ ($\nu=0.04721/\tauai$), that are identical to those obtained from the power spectra in the first iteration. All other normal modes are virtually absent in the numerical simulations of the last iteration.

The iterative method yields another approximation to the frequency that comes from the CEOF analysis of the numerical simulation of iteration \#6. A linear least-squares fit to the temporal phase gives the frequency $\omega=0.1016/\tauai$ for the fundamental evanescent mode and $\omega=0.2969/\tauai$ for the second longitudinal evanescent overtone. These values are identical to those of the first iteration. The error associated to these approximate values is 0.5\% and 0.7\%, so that the obtained accuracy is excellent.


\subsection{Error and stopping criterion}
\label{sec:iterative_error}

\begin{figure}[ht!]
  \centering{\includegraphics[width=\textwidth,angle=0]{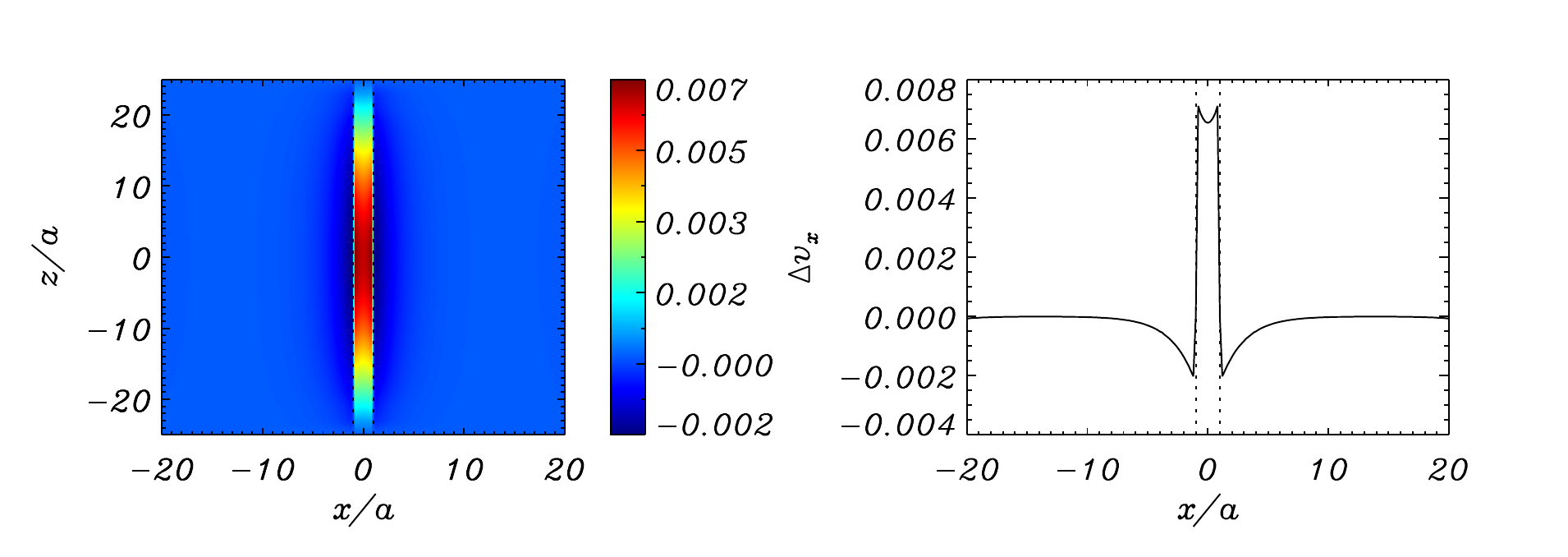}}
  \centering{\includegraphics[width=\textwidth,angle=0]{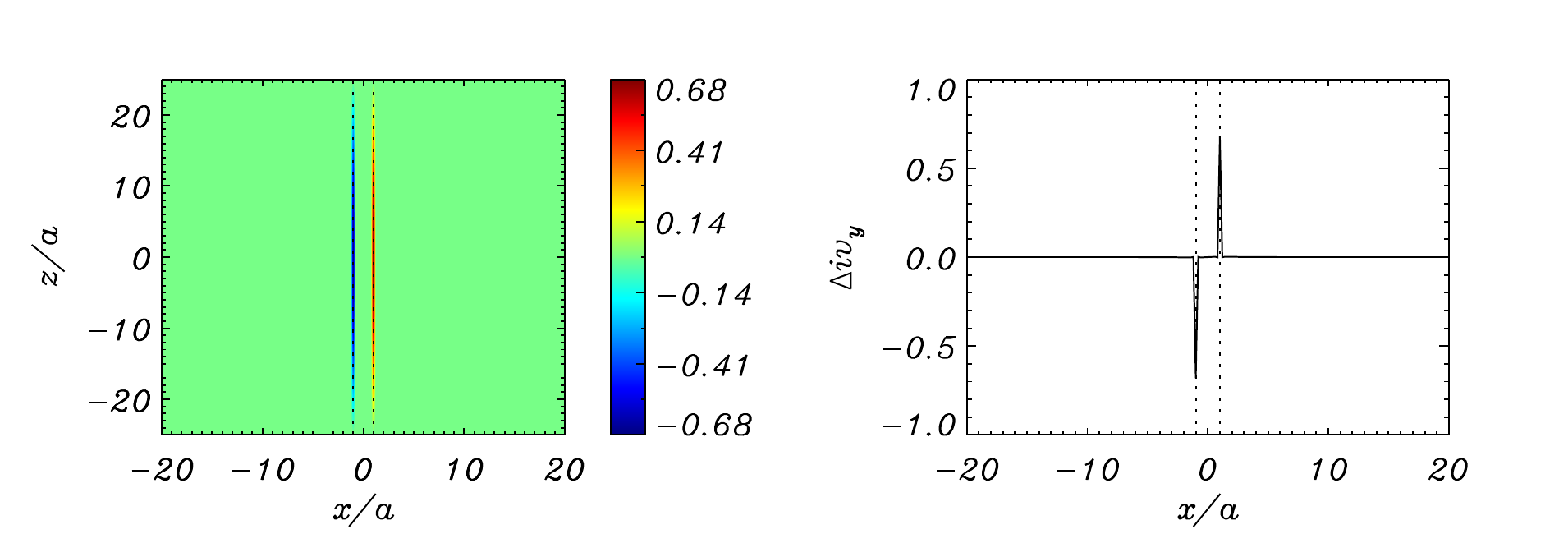}}
  \centering{\includegraphics[width=\textwidth,angle=0]{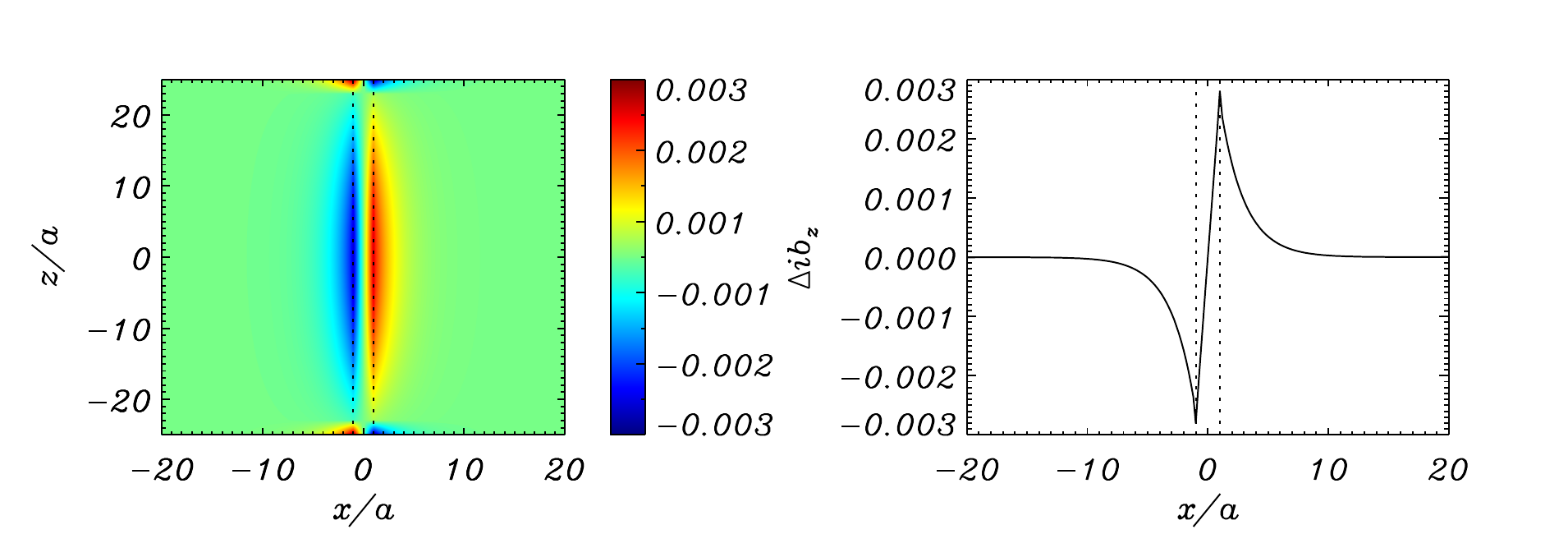}}
  \caption{Same as Figure~\ref{fig:ceof_vxvybz_comparison} for the CEOF analysis of iteration \#6.}
  \label{fig:ceof_vxvybz_comparison_iter4}
\end{figure}

\begin{figure}[ht!]
  \centering{\includegraphics[width=\textwidth,angle=0]{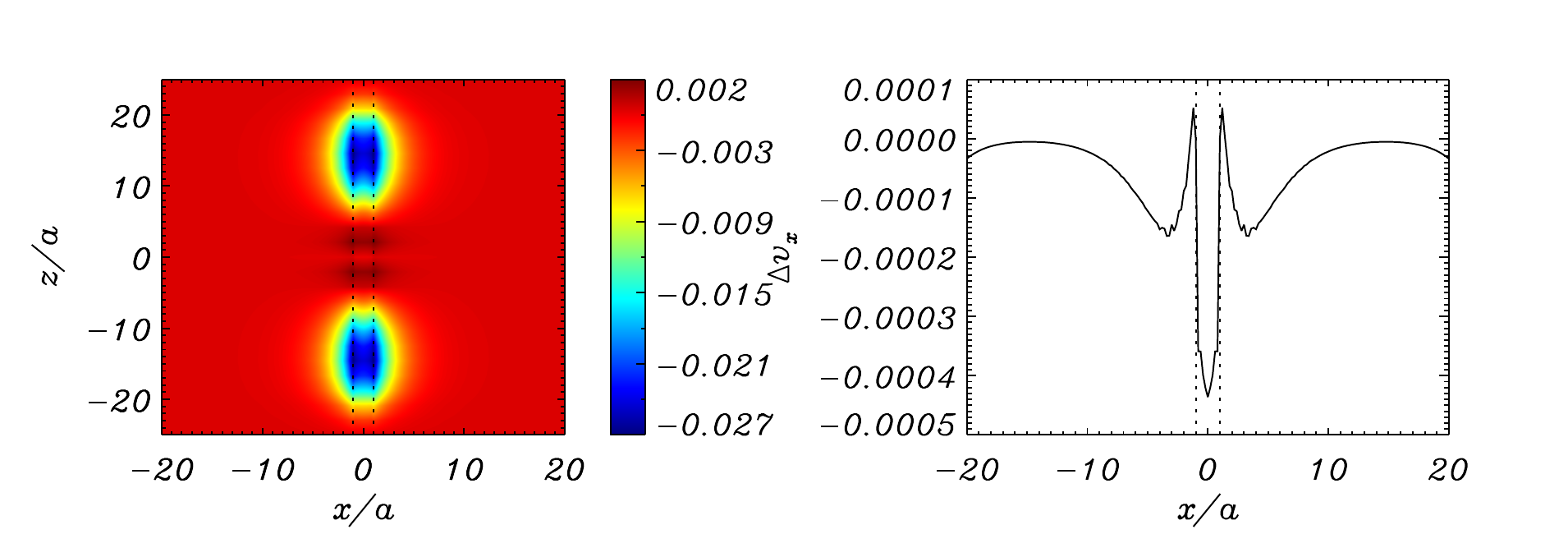}}
  \centering{\includegraphics[width=\textwidth,angle=0]{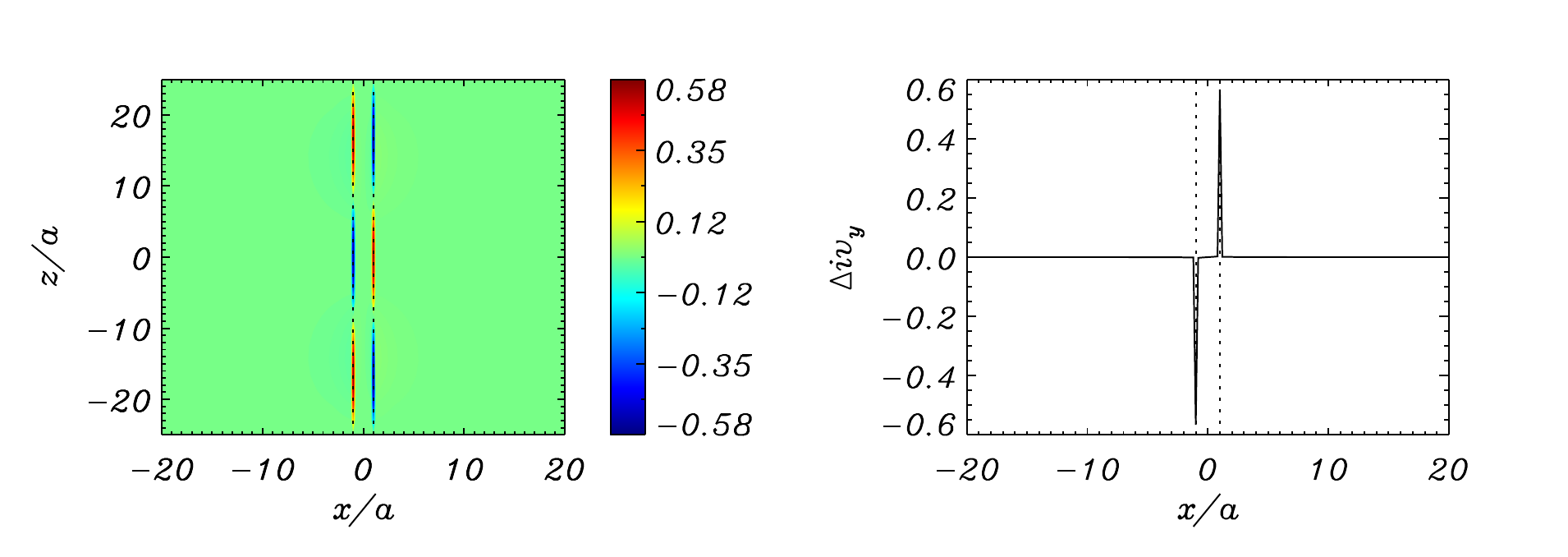}}
  \centering{\includegraphics[width=\textwidth,angle=0]{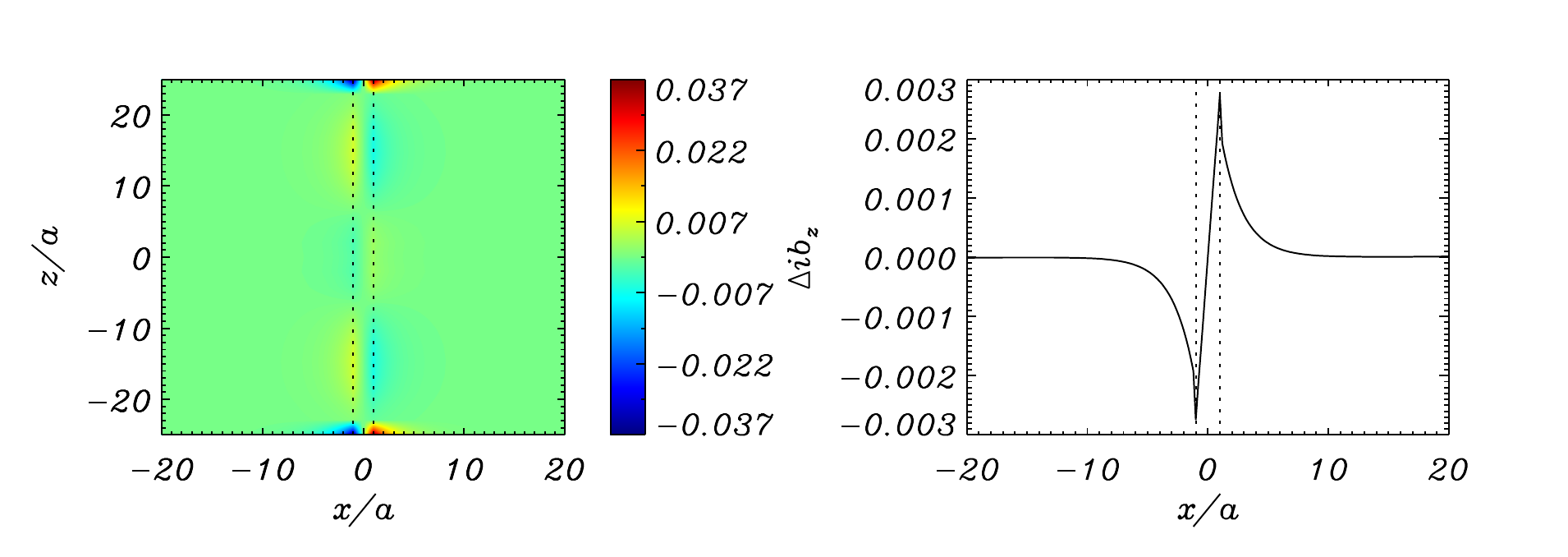}}
  \caption{Same as Figure~\ref{fig:ceof_vxvybz_comparison_2nd} for the CEOF analysis of iteration \#6.}
  \label{fig:ceof_vxvybz_comparison_iter4_2nd}
\end{figure}

We next examine in detail the error of the CEOF approximation to the eigenfunctions. Again we consider the results of iteration \#6 and show these errors in Figures~\ref{fig:ceof_vxvybz_comparison_iter4} and \ref{fig:ceof_vxvybz_comparison_iter4_2nd}. Although their maximum values are reduced by a factor of 2, the errors display some of the patterns of iteration~\#1: the error of $\vx$ is maximum in the slab, the error of $i\vy$ has a dominant component at the slab boundaries, but has been strongly reduced inside the slab during the iterative process, and the error of $\bz$ is maximum at the slab boundaries, but has become much more confined to the slab neighborhood.

We finally analyze the evolution of the error with the iterations\footnote{Before computing the errors described here we normalyze the normal mode eigenfunctions and their approximation from the CEOF analysis so that $\vx$ equals zero at $x=0$, $z=0$.}. At the end of iteration~\#$n$, with $n=1,2,\ldots$, Equations~(\ref{eq:ceof_mode_compare_vx}), (\ref{eq:ceof_mode_compare_vy}), and (\ref{eq:ceof_mode_compare_bz}) provide us with approximations for the three main eigenfunctions; we denote these approximations with the superscript $n$, i.e., $\vxceof^n$, $i\vyceof^n$, $i\bzceof^n$. For each eigenfunction, we define a global measure of the error, $\varepsilon$, by summing over the spatial domain the squares of the difference between the normal mode eigenfunction and the CEOF approximation. For example, for the iteration \#$n$ and the variable $\vx$ this global error is:

\begin{equation}\label{eq:global_error_vx}
\epsvx^n = \frac{1}{\Nx\Nz\max_{i,k}{|\hvx(x_i)\cos k_zz_k|}}\left\{\sum_{i,k}\left[\hvx(x_i)\cos k_zz_k-\vxceof^n(x_i,z_k)\right]^2\right\}^{1/2}, \quad n=1,2,\ldots,
\end{equation}

\noindent where the factor $\max_{i,k}{|\hvx(x_i)\cos k_zz_k|}$ in the denominator provides the right normalization that enables us to compare the error of different eigenfunctions. The additional factors $\Nx$ and $\Nz$ give an additional normalization that removes the dependence of $\epsvx^n$ on the number of points in the CEOF analysis. The definitions of $\epsvy^n$ and $\epsbz^n$ are done in a similar way.

\begin{figure}[ht!]
  \centering{\includegraphics[width=0.49\textwidth,angle=0]{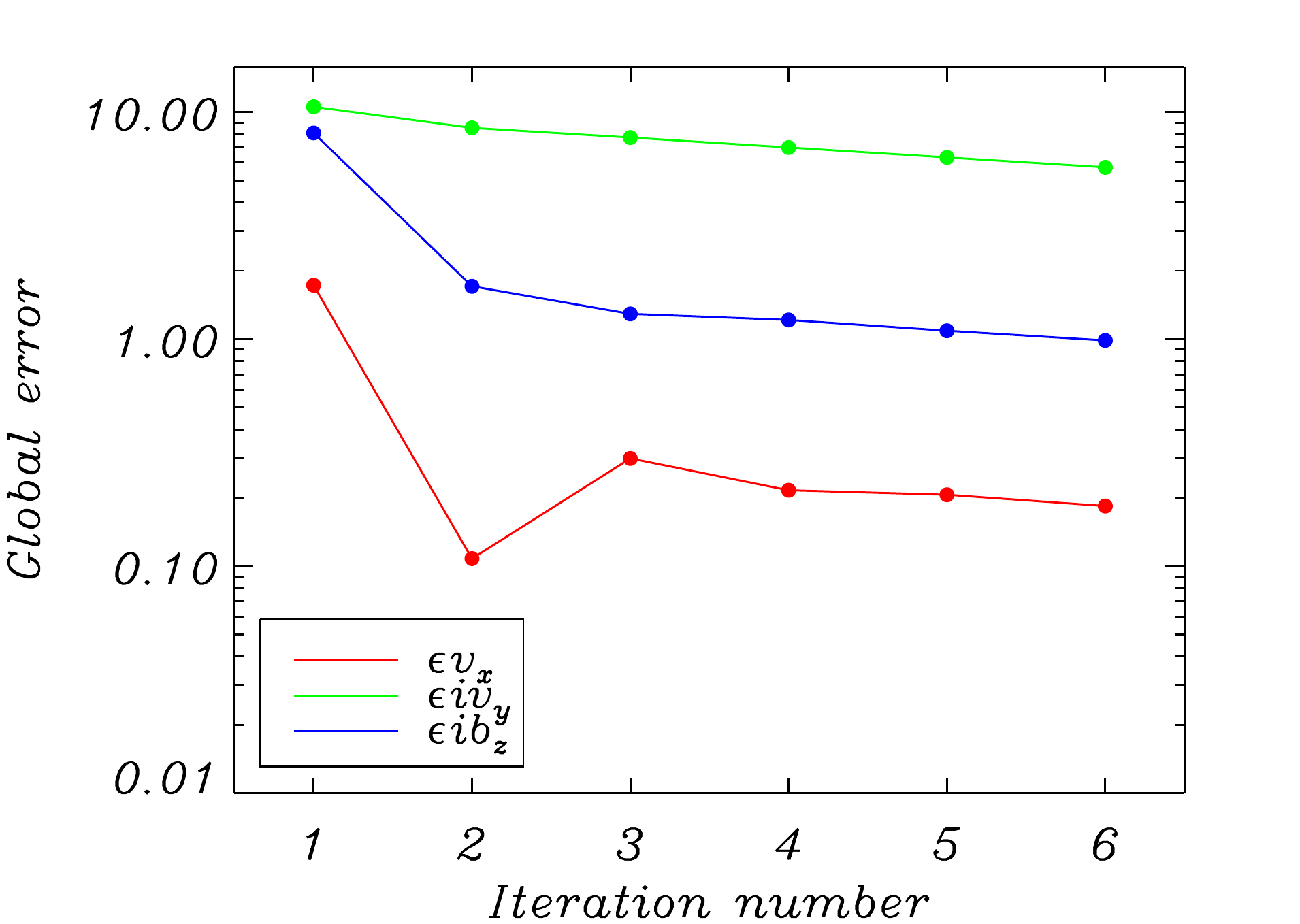} \includegraphics[width=0.49\textwidth,angle=0]{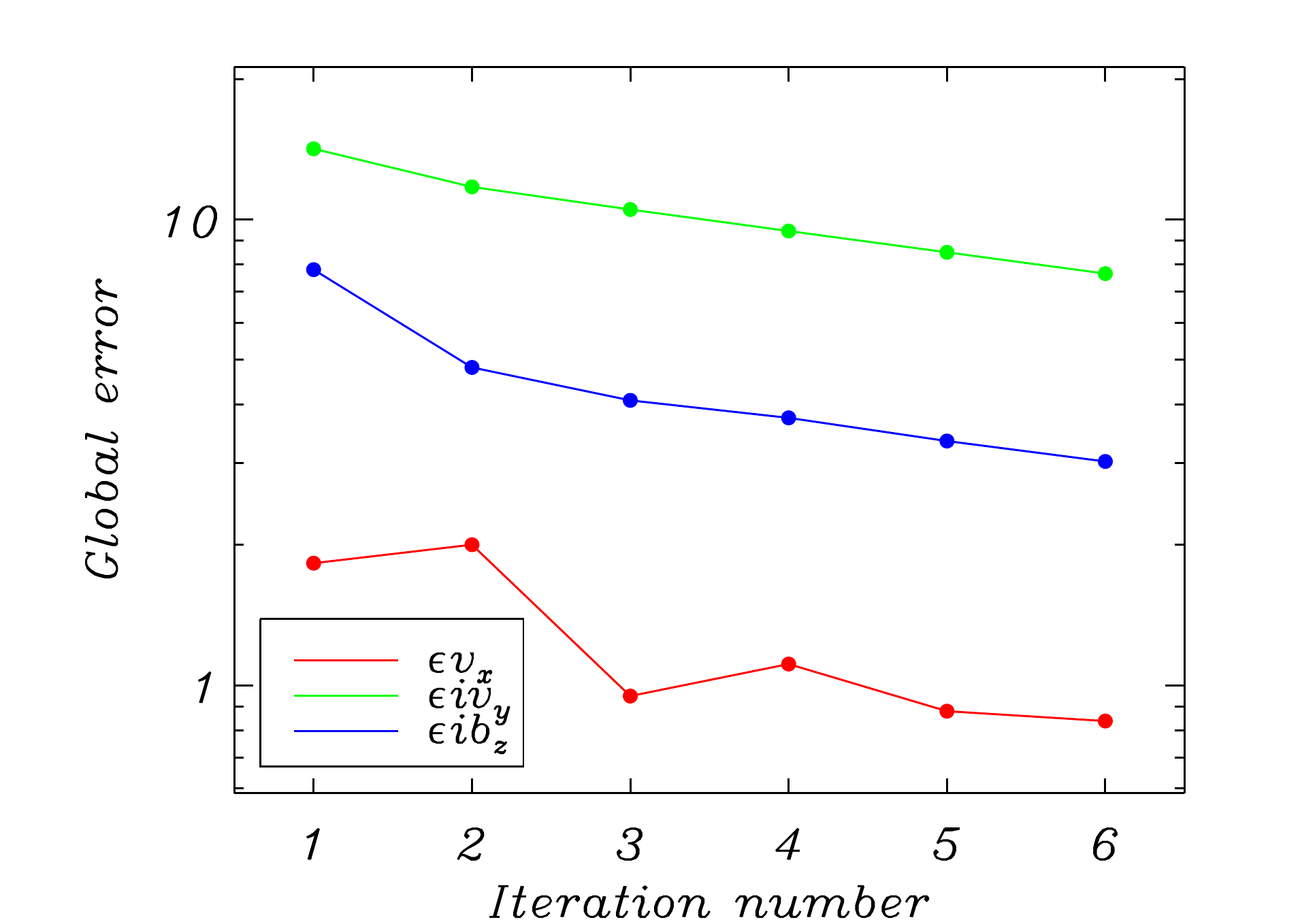}}
  \centering{\includegraphics[width=0.49\textwidth,angle=0]{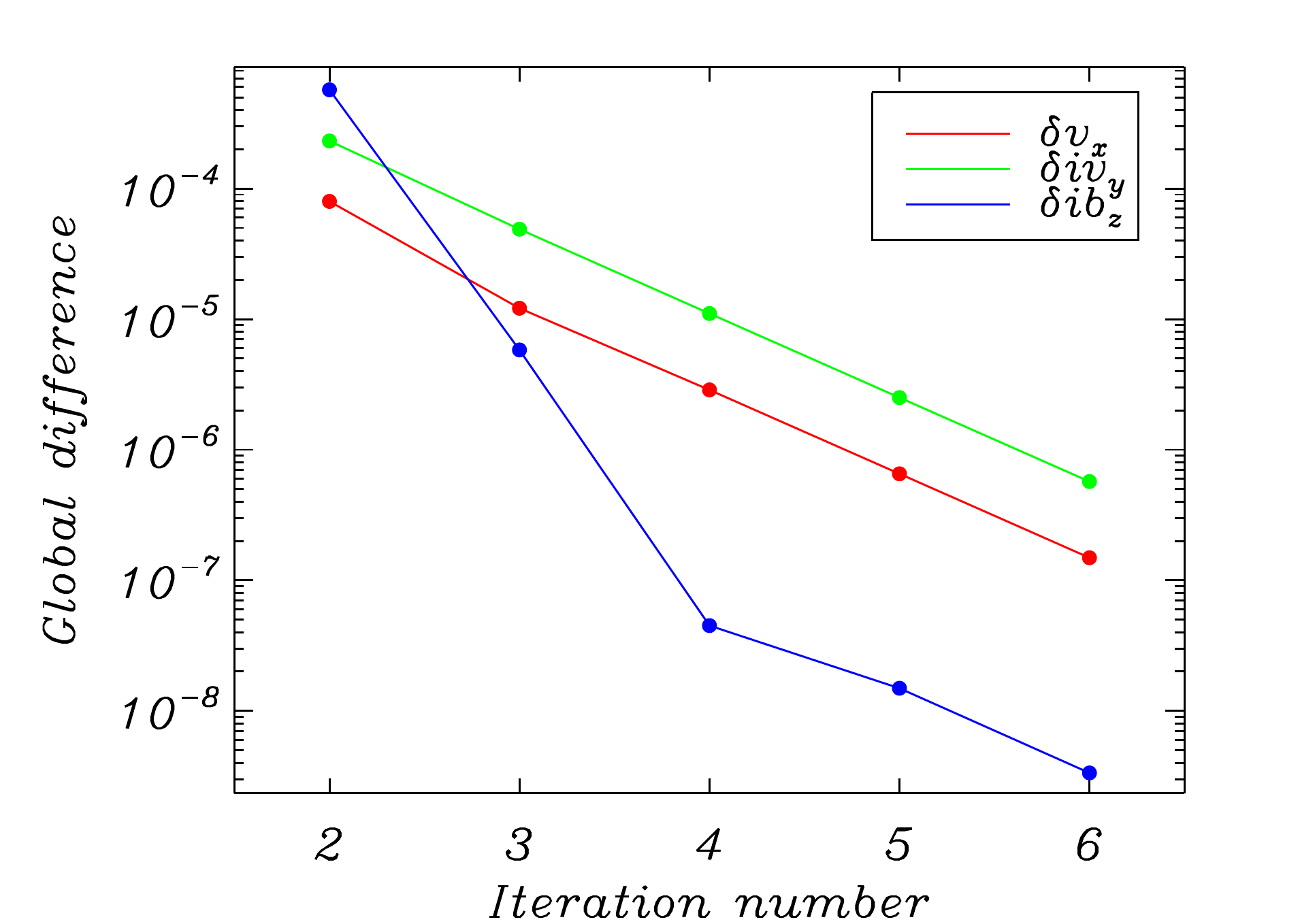} \includegraphics[width=0.49\textwidth,angle=0]{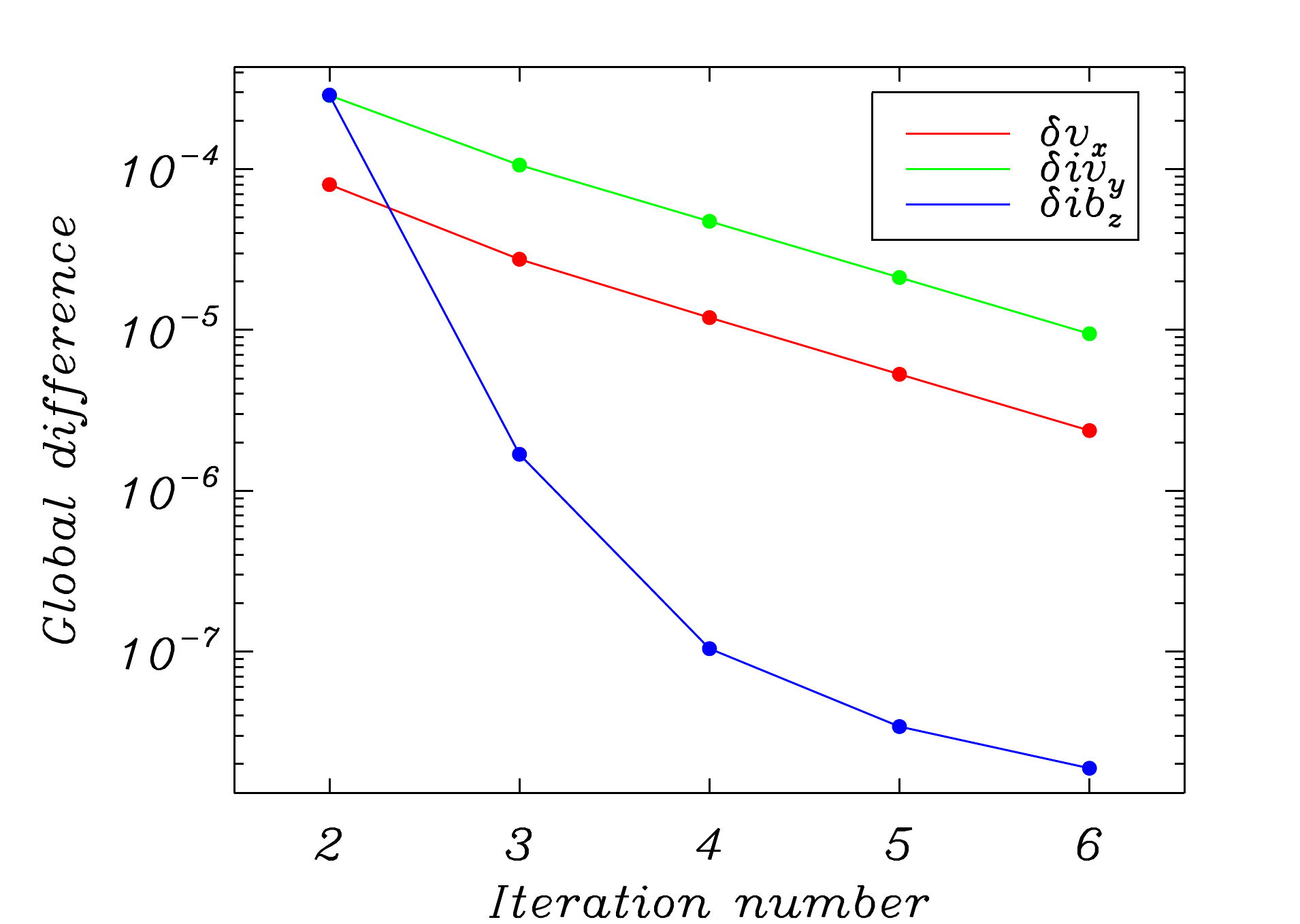}}
  \caption{Global error (top panels) and global difference (bottom panels) as a function of the iteration number. Left:~fundamental evanescent mode, right: second longitudinal evanescent overtone.}
  \label{fig:errors}
\end{figure}

The top panels of Figure~\ref{fig:errors} present the global errors for the first 6 iterations for the fundamental evanescent mode (left column) and its second longitudinal overtone (right column). In each iteration, $\vx$ has the smallest error (possibly because it is the eigenfunction with less ``contamination'' from the confined normal mode) and $i\vy$ displays the largest global error (because of the large contributions at the slab boundaries, that do not disappear with the iterative process). We see that the biggest improvement in the global error is obtained in iteration \#2, for which a remarkable reduction in $\epsvx$ and $\epsbz$ is found. The subsequent variation of the three global errors is much more moderate and so for this problem two iterations give a good compromise between the error associated to the CEOF approximations and the computer time spent.

The case studied in this paper allows us to compute the global error because of our knowledge of the exact eigenfunctions. In a general case, in which the eigenfunctions are unknown and our aim is just to obtain them, a proxy for the global error can be computed by comparing the approximate eigenfunctions of two successive iterations. To define this new global uncertainty measure, in Equation~(\ref{eq:global_error_vx}) we replace the normal mode eigenfunction $\hvx(x_i)\cos k_zz_k$ by its CEOF approximation in the iteration $n+1$. We also rewrite the iteration indices and substitute $n+1$ by $n$. This gives the following definition for the global difference between the $\vx$ eigenfunction of iterations \#$n$ and \#$n-1$:

\begin{equation}\label{eq:global_difference_vx}
\difvx^{n} = \frac{1}{\Nx\Nz\max_{i,k}{|\vxceof^{n}(x_i,z_k)|}}\left\{\sum_{i,k}\left[\vxceof^{n}(x_i,z_k)-\vxceof^{n-1}(x_i,z_k)\right]^2\right\}^{1/2}, \quad n=2,3,\ldots.
\end{equation}

\noindent Analogous expressions can be written for $\difvy$ and $\difbz$.

The variation of the global difference with the iterations is displayed in the bottom panels of Figure~\ref{fig:errors}. The convergence is quite fast, with $\difbz$ showing an improvement of roughly two orders of magnitude per iteration during the first iterations, the highest convergence rate of all variables. $\difvx$ and $\difvy$ are reduced at a slower pace, namely, almost an order of magnitude per iteration. We see that all variables attain a global difference smaller than $10^{-5}$ in iteration \#5 (fundamental evanescent mode) and in iteration \#6 (second longitudinal evanescent overtone). Therefore, we adopt the iteration stopping criterion that the global difference must be smaller than $10^{-5}$.

\section{CONCLUSIONS}
\label{sec:conclusions}

In this paper we have devised a method to determine a physical system normal modes, i.e., their eigenfunctions and eigenfrequencies, by the iterative application of time-dependent numerical simulations of the equations that govern the system dynamics and the CEOF analysis of the simulation results. We have illustrated how the CEOF method can be applied to all the non-redundant variables: in our case, in particular, this means that we can avoid including $\bx$ and $i\by$ in the CEOF computation because their eigenfunctions can be readily computed from the other three ($\vx$, $i\vy$, and $\bz$). At the end of each iteration, the CEOF approximations to the eigenfunctions are used as the initial conditions for the time-dependent numerical simulation of the next iteration. Finally, we have examined the global error of the approximate eigenfunctions as a function of the iteration number and have established a convergence criterion based on the global difference between the approximate eigenfunctions of consecutive time steps.


The main disadvantage of our test case is the presence of sharp boundaries in the equilibrium structure, which leads to abrupt jumps of the eigenfunction $i\hvy$ and non-derivable eigenfunctions $\hvx$ and $i\hbz$ at these positions (see Figure~\ref{fig:kink_eigenfunctions}). We have found that these normal mode features result in the presence of large errors at the slab boundary, which are quite substantial for the approximation to $i\hvy$; see Figures~\ref{fig:ceof_vxvybz_comparison_iter4} and \ref{fig:ceof_vxvybz_comparison_iter4_2nd}.

We have obtained an approximation to the two normal modes of interest (the fundamental evanescent mode and its second longitudinal overtone) with great accuracy: after 6 iterations, the frequency is wrong by only 0.5\%--0.7\% and the eigenfunctions $\hvx$ and $i\hbz$ have maximum errors of the order of 0.6\% and 0.7\%, respectively. The case of $i\hvy$ is worse because of the difficulties of recovering a function that jumps at $x=\pm a$. If these two lines are ignored, the maximum error of $i\hvy$ is also of the order of 0.6\%.



In the second paper of this work \citep{rial2019} we will apply the technique presented here to the time-dependent numerical simulations of a loop embedded in a coronal arcade carried out by \citet{rial2013}. The equilibrium structure is similar to the one used in the present paper but includes a curved slab in which both the magnetic field strength and plasma density vary along the magnetic field. The initial condition used by \citet{rial2013} is analogous to Equation~(\ref{eq:initial_disturbance}) and so various longitudinal harmonics are excited. The present paper shows that our technique is well suited for this task because it allows to obtain the features of different normal modes. Its application to the more realistic numerical simulations by \citet{rial2013} should produce normal mode characteristics comparable to observed loop oscillation events.

\section*{Acknowledgements}
R.O. and J.T. acknowledge support from the Spanish Ministry of Economy and Competitiveness (MINECO) and FEDER funds through project AYA2017-85465-P. I.A. acknowledges support by the Spanish Ministry of Economy and Competitiveness (MINECO) through projects AYA2014-55456-P (Bayesian Analysis of the Solar Corona) and AYA2014-60476-P (Solar Magnetometry in the Era of Large Solar Telescopes) and FEDER funds. R.O. is indebted to D. W. Fanning for making available the Coyote Library of IDL programs (http://www.idlcoyote.com/).

\bibliography{references}
\bibliographystyle{apj}

\end{document}